\documentclass[aps,twocolumn,pra,superscriptaddress,notitlepage]{revtex4-2}
\usepackage{graphicx}
\usepackage{dcolumn}
\usepackage{bm}
\usepackage[breaklinks=true,colorlinks=true,linkcolor=blue,urlcolor=blue,citecolor=blue]{hyperref}
\usepackage{physics}
\usepackage{physics}

\begin{document}

\title{Generating large-scale Greenberger-Horne-Zeilinger-like states in lattice spin systems}
\author{Xuanchen Zhang}
\thanks{These authors contributed equally.}
\affiliation{State Key Laboratory of Low-Dimensional Quantum Physics, Department of
Physics, Tsinghua University, Beijing 100084, China }
\author{Yaofeng Chen}
\thanks{These authors contributed equally.}
\affiliation{State Key Laboratory of Low-Dimensional Quantum Physics, Department of
Physics, Tsinghua University, Beijing 100084, China }
\author{Yong-Chun Liu}
\email{ycliu@tsinghua.edu.cn}
\affiliation{State Key Laboratory of Low-Dimensional Quantum Physics, Department of
Physics, Tsinghua University, Beijing 100084, China }
\affiliation{Frontier Science Center for Quantum Information, Beijing 100084, China}
\date{\today}

\begin{abstract}
Greenberger-Horne-Zeilinger (GHZ) state is a typical maximally entangled 
state which is pursued in both fundamental research and emerging quantum
technologies. Preparing large-scale GHZ states in lattice spin systems is
particularly appealing for quantum advantages, but conventional schemes face
great challenges in scalability. Here we propose a universal and scalable
scheme to generate large-scale GHZ-like states, which share similar
entanglement and metrological properties with standard GHZ states, in
lattice spin systems through global Floquet engineering. Our scheme requires
only global operations and shows great advantage for large particle number.
It is applicable to systems with arbitrary interaction ranges, offering a
practical pathway for large-scale implementation of many-body entangled
states in various systems.
\end{abstract}

\maketitle

\section{Introduction}\label{sec:1}

Lattice spin systems have emerged as promising
platforms for quantum computation \cite{Arute2019, Evered2023, Moses2023},
quantum simulation \cite{Scholl2021, Cornish2024, Guo2024}, and quantum
metrology \cite{Matthew2019, Madjarov2019, Eckner2023}, owing to their
advantages in addressability and control. Central to these applications is
the ability to create large-scale nonclassical states with many-body
entanglement, which is a key resource in the pursuit of quantum advantages
\cite{Giovannetti2004, Cao2023}. A typical example is scalable spin-squeezed
states \cite{Kitagawa1993, Wineland1994, Esteve2008, Jin2009, Ma2011,
Liu2011, Chen2019, Bao2020, Huang2021, Huang2023, Hu2023}, which enable
quantum-enhanced metrology \cite{Gross2010, Riedel2010, Pezze2018,
Greve2022, Mao2023, Robinson2024} and have recently been widely investigated
in lattice spin platforms \cite{Michael2016, Bohnet2016, Perlin2020,
Comparin2022, Young2023, Bornet2023, Franke2023, Block2024}. A more
ambitious target is the Greenberger-Horne-Zeilinger (GHZ) state \cite{Greenberger1990}, 
which is a representative example of maximally entangled state. They are
not only crucial in quantum technologies such as quantum precision
measurement \cite{Bollinger1996, Leibfried2004}, but also play vital roles
in fundamental researches covering macroscopic quantum effects \cite%
{Frowis2018} and quantum nonlocality \cite{Greenberger1990, Mermin1990}.
Experimentally, GHZ states have been realized across various systems,
including superconducting circuits \cite{Song2017, Song2019, Bao2024},
trapped ions \cite{Monz2011, Choi2014}, Rydberg atom arrays \cite{Omran2019}%
, and photonic systems \cite{Wang2016, Wang2018}.

However, previous schemes for generating GHZ states encounter significant
challenges in scalability. Typically, GHZ states are prepared either through
sequences of two-qubit gates \cite{Kaufmann2017, Wei2020, Mooney2021} or via
optimal control involving evolution in full Hilbert space \cite{Lu2019, Omran2019, Cao2024}.
These approaches have limitations for large particle number due to the
accumulation of errors in gate operation and increasement of complexity in
optimal control. These constraints significantly impede the broader
exploitation of GHZ states in fundamental quantum researches and emerging
quantum technologies.

In this work, we present a universal and scalable scheme to generate
large-scale GHZ-like states in lattice spin systems via global Floquet
engineering. The generated GHZ-like states are similar to standard GHZ
states in entanglement and metrological properties. Considering the
power-law Ising model, we find that the effective interaction induced by a periodic
sequence of global single-qubit rotations includes a collective-spin three-body term, which can create GHZ-like states for large particle number. The spin-wave excitations can be
suppressed by reducing the separation between the rotation pulses, making our scheme applicable to extensive interaction ranges. We further take into account typical incoherent noises and demonstrate the robustness of our scheme against decoherence.

\section{System model and our scheme}\label{sec:2}

We consider the power-law Ising model with the interaction described by
\begin{eqnarray}\label{H-Ising}
H=\sum_{j\neq k}K_{jk}s_{j}^{z}s_{k}^{z},
\end{eqnarray}
where $s_{j}^{z}=\sigma _{j}^{z}/2$ is the $z$ component of the $j$-th spin,
and the coupling strength between spins $j$ and $k$ located at positions $\bm{r}_{j}$ and $\bm{r}_{k}$ is given by
\begin{eqnarray}
     K_{jk}=K/\abs{\bm{r}_j-\bm{r}_k}^{\alpha },
\end{eqnarray}
with $K$ being the interaction strength of the nearest-neighbor
pair and $\alpha $ being the decaying factor. In this paper, we
focus on the cases of 1-dimensional (1D) and 2-dimensional (2D) square lattices.
Such a model can be realized in a variety of lattice spin systems including trapped ions ($0<\alpha <3$)
\cite{Britton2012, Monroe2021}, Rydberg atom arrays ($\alpha =3,6$) \cite%
{Schauss2012, Browaeys2020}, polar molecules ($\alpha =3$) \cite{Miller2024}%
, superconducting circuits ($\alpha \rightarrow \infty $, the limit of
nearest-neighbor interactions) \cite{Philipp2022}, and others. Note that the
Hamiltonian \eqref{H-Ising} reduces to the well-known one-axis twisting
(OAT) model \cite{Kitagawa1993}
\begin{eqnarray}
     H_{\mathrm{OAT}}=K(S_{z}^{2}-N/2),     
\end{eqnarray}
when interactions are uniform ($\alpha =0$), which are often referred to as
all-to-all or infinite-range interactions, where $S_{z}=\sum_{i=1}^{N}s_{i}^{z}$ denotes the $z$ component of the collective spin
operator $\bm{S}$ with total spin $S=N/2$, and $N$ is the total particle
number.

\begin{figure}[tbp]
\includegraphics[width=\columnwidth]{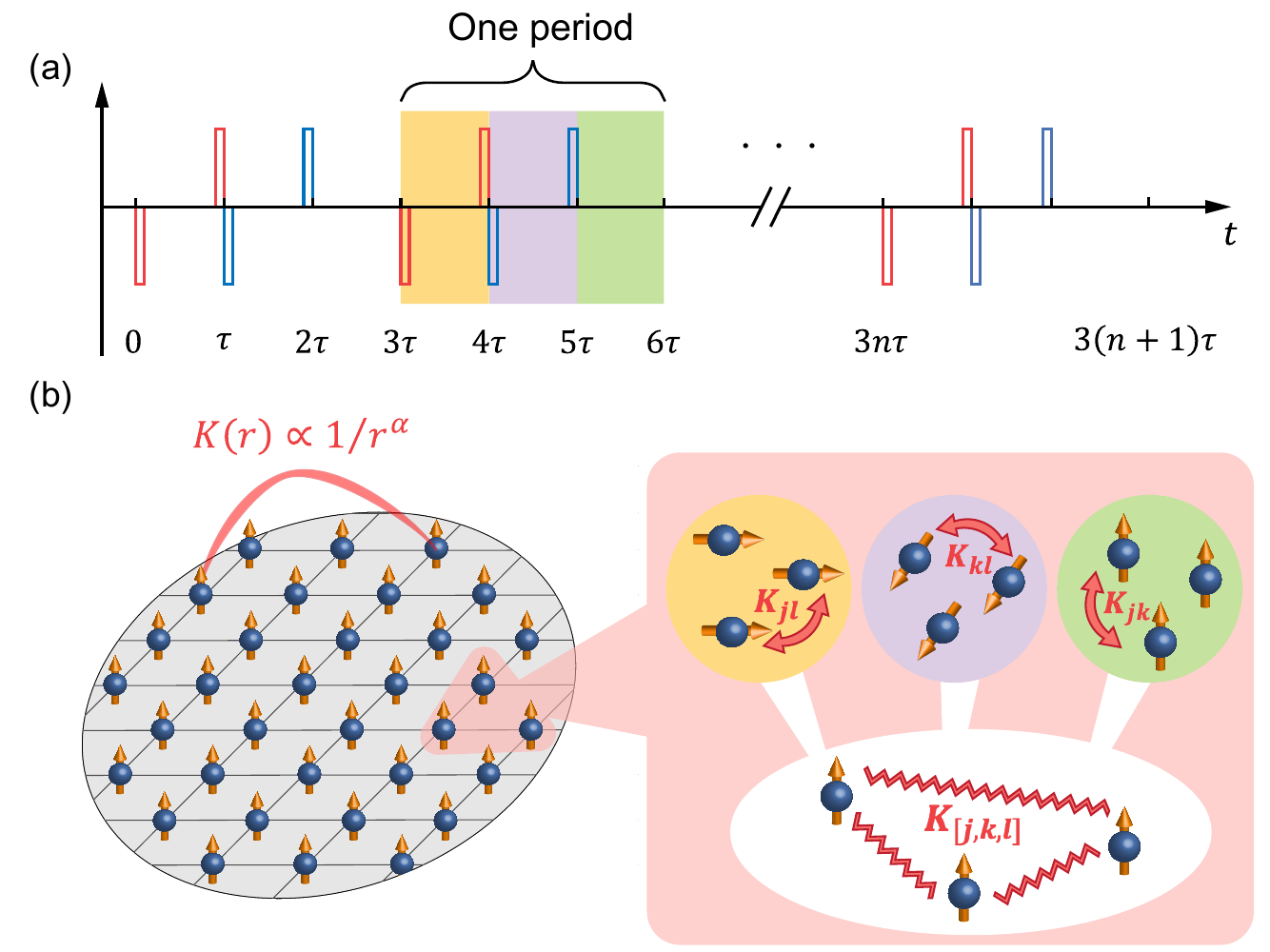}
\caption{Schematic diagram of the scheme to generate GHZ-like states in
lattice spin systems governed by the power-law Ising model. (a) An
illustration of the pulse sequences. The red and blue pulses denote $\pm%
\protect\pi/2$ pulses along $x$ and $y$ axis, respectively. The spins
effectively interact through $s_j^y s_k^y$ (yellow shaded), $s_j^x s_k^x$
(purple shaded) and $s_j^z s_k^z$ (green shaded) in the three successive
parts of one period, each taking the time of $\protect\tau$. (b) An
illustration of the model. The spins located in a square lattice interact
through power-law Ising interaction, with interaction strength decaying with
distance $r$ as $1/r^\protect\alpha$. The noncommutativity of the three
segments in one period gives rise to the effective three-body interaction. }
\label{fig:1}
\end{figure}

To generate GHZ-like states, we introduce the global Floquet-driving scheme involving iterative $\pi /2$ pulses along $x$- and $y$-axes \cite{Zhang2024}, as shown in Fig.~\ref{fig:1}(a). The evolution operator over one period is given by
\begin{eqnarray}\label{Uoperator}
     U(3\tau) &=& e^{-i H\tau} e^{-i\frac{\pi}{2}S_y} e^{-i H\tau} e^{i\frac{\pi}{2}S_y} e^{-i\frac{\pi}{2}S_x} e^{-i H\tau} e^{i\frac{\pi}{2}S_x} \nonumber \\ 
     &=& e^{-i H_{zz}\tau} e^{-i H_{xx}\tau} e^{-i H_{yy}\tau},
\end{eqnarray}
where $H_{\mu\mu} = \sum_{j \neq k} K_{jk} s_j^\mu s_k^\mu$ ($\mu = x,y,z$), and $\tau$ denotes the separation between two $\pm \pi /2$ pulses. This pulse sequence effectively engineers spin-spin interactions of the forms $s_{j}^{y}s_{k}^{y}$, $s_{j}^{x}s_{k}^{x}$ and $s_{j}^{z}s_{k}^{z}$ in the three segments of each cycle, respectively. For sufficiently small $\tau$, the evolution operator can be approximated as 
\begin{eqnarray}
     U(3\tau)=\exp\left\{-i3\tau H_{\mathrm{eff}} + \mathcal{O}(\tau^2)\right\},
\end{eqnarray}
where the effective Hamiltonian is obtained using the Baker-Campbell-Hausdorff expansion and takes the form (see Appendix \ref{app:1} for detailed derivations)
\begin{eqnarray}
H_{\mathrm{eff}} &=&\frac{1}{3}\sum_{j\neq k}K_{jk}\bm{s}_{j}\cdot \bm{s}_{k}
\nonumber  \label{H-eff} \\
&&+\frac{1}{3}\tau \sum_{\lbrack
j,k,l\rbrack}K_{[j,k,l]}(s_{j}^{x}s_{k}^{y}s_{l}^{z}+s_{l}^{z}s_{k}^{y}s_{j}^{x}),
\end{eqnarray}
where $[j,k,l]$ stands for a combination of mutually distinct indices $j,k,l$, and the effective three-body interaction strength is given by
\begin{eqnarray}
     K_{[j,k,l]}=K_{jl}K_{kl}+K_{jk}K_{lk}-K_{kj}K_{lj}.
\end{eqnarray}
The requirement for safely ignoring higher order terms $\mathcal{O}(\tau^2)$ imposes the condition $\tau\ll\tau_{\mathrm{crit}}\sim L^{-\mathrm{max}(0, d-\alpha)}$ for $d$-dimensional square lattice with lattice length $L$.

The effective Hamiltonian \eqref{H-eff} consists of two terms with clear physical interpretations. The zeroth-order term corresponds to the Heisenberg coupling $\bm{s}_{j}\cdot \bm{s}_{k}=s_{j}^{x}s_{k}^{x}+s_{j}^{y}s_{k}^{y}+s_{j}^{z}s_{k}^{z}$, which
induces a many-body gap between the permutationally symmetric manifold (the
Dicke manifold) and the others, thus suppresses the leakage of population
outside the Dicke manifold. The first-order term is the effective three-body
interaction originated from the noncommutativity of three segments within one Floquet
period, as shown in Fig.~\ref{fig:1}(b). Crucially, the protection provided by the Heisenberg coupling effectively retains the collective
part of this term, which is the key mechanism enabling the generation of GHZ-like states in our scheme.

\section{Zero-momentum/finite-momentum decomposition}\label{sec:3}

In this section, we study the effective Hamiltonian \eqref{H-eff} by applying the zero-momentum/finite-momentum decomposition \cite{Roscilde2023}. This treatment serves to demonstrate explicitly that our scheme can generate GHZ-like states and to identify the corresponding parameter regime.

We begin by introducing the Holstein-Primakoff (HP) transformation
\begin{eqnarray}\label{HP-trans}
     s_j^x &=& \frac{1}{2}\left(\sqrt{1-b_j^\dagger b_j}b_j + b_j^\dagger \sqrt{1-b_j^\dagger b_j}\right), \nonumber \\
	s_j^y &=& \frac{1}{2i}\left(\sqrt{1-b_j^\dagger b_j}b_j - b_j^\dagger \sqrt{1-b_j^\dagger b_j}\right), \nonumber \\
	s_j^z &=& \frac{1}{2} - b_j^\dagger b_j.
\end{eqnarray}
Exploiting the translational symmetry of the system, it is convenient to define the HP bosons in momentum space
\begin{eqnarray}
     b_{\bm{q}}=\sum_j e^{-i\bm{q}\cdot\bm{r}_j}/\sqrt{N},
\end{eqnarray}
with momentum vectors $\bm{q} = 2\pi(q_1, q_2, ..., q_d)/L$, where $q_m = 0, 1, ..., L-1$.

For a generic operator $O$, we decompose it into zero- and finite-momentum components
\begin{eqnarray}
     O = [O]_{\mathrm{ZM}} + [O]_{\mathrm{FM}}.
\end{eqnarray}
The zero-momentum component contains only $q=0$ HP bosons, $[O]_{\mathrm{ZM}}=[O]_{\mathrm{ZM}}(b_0, b_0^{\dagger})$, which is exactly the permutationally symmetric part and can be calculated by projecting $O$ onto the Dicke manifold $\{\ket{S=N/2,M}\equiv\ket{M}\}$ \cite{Roscilde2023}:
\begin{eqnarray}
    [O]_{\mathrm{ZM}} = \sum_{M,M'}\mel{M}{O}{M'}\ket{M}\bra{M'}.
\end{eqnarray}
By contrast, the finite-momentum component $[O]_{\mathrm{FM}}$ describes the non-collective fluctuations. Provided these fluctuations remain small during the whole evolution, they can be treated perturbatively by retaining up to quadratic terms of the bosonic operators $b_{\bm{q}}^2$, $b_{\bm{q}}^{\dagger 2}$ and $b_{\bm{q}}^{\dagger}b_{\bm{q}}$.

Applying this decomposition to the effective Hamiltonian \eqref{H-eff}, we approximately split it into two independent parts (see Appendix \ref{app:2}):
\begin{eqnarray}\label{ZM-FM}
     H_{\mathrm{eff}}\simeq H_{\mathrm{ZM}}+H_{\mathrm{SW}}.
\end{eqnarray}
The zero-momentum component
takes exactly the form of the collective-spin cubic XYZ model as defined in \cite{Zhang2024}:
\begin{eqnarray}\label{H-ZM}
     H_{\mathrm{ZM}}=\frac{1}{3}\lambda K^{2}\tau \left(
     J_{x}J_{y}J_{z}+J_{z}J_{y}J_{x}\right),
\end{eqnarray}
where $J_{\mu }=\sum_{M,M^{\prime }}\ket{M}\bra{M}S_{\mu }\ket{M'}\bra{M'}$ is the
zero-momentum component of the collective-spin operator $S_{\mu}$, which has a fixed spin length $J=N/2$. This Hamiltonian governs the purely collective dynamics within the Dicke manifold and is responsible for the generation of GHZ-like states. The finite-momentum component, up to quadratic terms of the HP bosons, corresponds to the spin-wave Hamiltonian
\begin{eqnarray}
     H_{\mathrm{SW}} &=&-\frac{N}{12}K_{0}+\frac{1}{3}\sum_{\bm{q}\neq 0}\left(
     K_{0}-K_{\bm{q}}\right) b_{\bm{q}}^{\dagger }b_{\bm{q}}  \nonumber
     \label{H-SW} \\
     &&-i\frac{\tau }{12}\sum_{\bm{q}\neq 0}\left( K_{\bm{q}}^{2}-T_{0}^{2}%
     \right) \left( b_{\bm{q}}b_{-\bm{q}}-b_{\bm{q}}^{\dagger }b_{-\bm{q}%
     }^{\dagger }\right),
\end{eqnarray}
where the coefficients are defined as
\begin{eqnarray}
     \lambda &=&\frac{1}{N(N-1)(N-2)}\sum_{[j,k,l]}\frac{1}{r_{jl}^{\alpha }}
     \frac{1}{r_{kl}^{\alpha }},  \nonumber \\
     K_{\bm{q}} &=&K\sum_{\bm{r}\neq 0}\frac{1}{r^{\alpha }}e^{-i\bm{q}\cdot\bm{r}},\quad T_{0}^{2}=K^{2}\sum_{\bm{r}\neq 0}\frac{1}{r^{2\alpha }}.
\end{eqnarray}
The undesired non-collective effects are described by the linear spin-wave excitations, which can be excatly solved via the Bogoliubov diagonalization
of $H_{\mathrm{SW}}$ \eqref{H-SW}, leading to (see Appendix \ref{app:2})
\begin{eqnarray}\label{SW-excit}
     \expval{b_{\bm{q}}^\dagger b_{\bm{q}}}(t)=\frac{\tau ^{2}}{2}\frac{B_{\bm{q}}^{2}}{A_{\bm{q}}^{2}-\tau ^{2}B_{\bm{q}}^{2}}\left[ 1-\cos \left( \epsilon_{\bm{q}}t\right) \right],
\end{eqnarray}
with $A_{\bm{q}}=(K_{0}-K_{\bm{q}})/3$, $B_{\bm{q}}=(K_{\bm{q}}^{2}-T_{0}^{2})/6$ and $\epsilon _{\bm{q}}=\sqrt{A_{\bm{q}}^{2}-\tau ^{2}B_{\bm{q}}^{2}}$.

The above decomposition neglects both the nonlinear behavior of the finite-momentum component and its coupling to the zero-momentum component. This approximation is reasonable as long as the finite-momentum excitations remain weak. From Eq.~\eqref{SW-excit}, we find that a sufficiently small pulse
separation $\tau \ll \mathrm{min}_{\bm{q}}[A_{\bm{q}}/B_{\bm{q}}]$ is able
to suppress the finite-momentum spin-wave excitations. Once this condition is fulfilled,
the evolution will be dominated by the zero-momentum component $H_{\mathrm{ZM}}$ \eqref{H-ZM}, generating a GHZ-like state from an initial coherent spin state \cite{Zhang2024}.

\section{Numerical investigation}

In this section, we investigate the performance of our scheme by numerically studying the time evolution governed by Eq.~\eqref{Uoperator}.

\subsection{Characterizing the GHZ-like state}

\begin{figure}[tbp]
\includegraphics[width=\columnwidth]{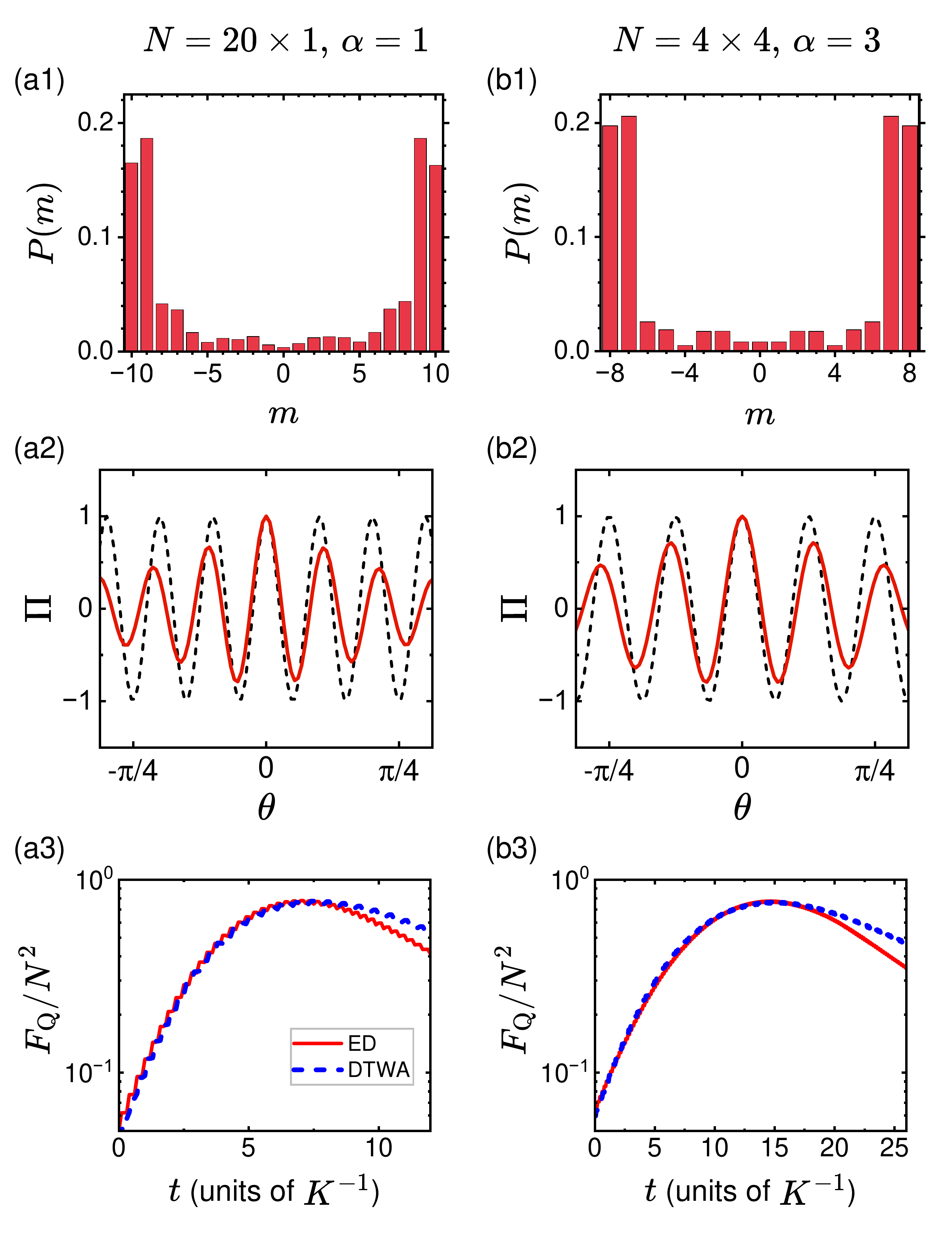}
\caption{Characterization of the GHZ-like state generated in our scheme, calculated for (a) $20\times1$ spins, $\alpha=1$, $K\tau=0.1$ and (b) $4\times4$ spins, $\alpha=3$, $K\tau=0.06$. (a1), (b1) The probability distributions $P(m)$ of the obtained states, where $m$ denotes the eigenvalue of $S_x$. (a2), (b2) Parity oscillations of the obtained states. Black dashed curves show the results of perfect GHZ states $\expval{\Pi(\theta)}_{\mathrm{GHZ}}=\cos N\theta$. (a3), (b3) Time evolution of the quantum Fisher information $F_\mathrm{Q}$, obtained using exact diagonalization (ED, red solid) and discrete truncated Wigner approximation (DTWA, blue dashed).}
\label{fig:2}
\end{figure}

Choosing the initial state as the coherent spin state polarized along $z$-axis $\ket{\psi(0)}=\ket{\uparrow}^{\otimes N}$, our scheme creates a GHZ-like state with probability distribution strongly concentrated near the extremal eigenvalues $S_x = \pm N/2$. As representative examples, we consider a 1-dimensional (1D) system of $20\times1$ spins with $\alpha=1$, and a 2D system of $4\times4$ spins with $\alpha=3$. The resulting probability distributions $P(m)\equiv P(S_x=m)$ of the obtained states are shown in Fig.~\ref{fig:2} (a1) and (b1). In both cases, the distributions are clearly bimodal, with most of the weight concentrated near $S_x = \pm N/2$, indicating the formation of collective superpositions.

The coherence of GHZ-like states can be further demonstrated by examining the parity operator
\begin{eqnarray}
     \Pi(\theta) = e^{-i S_x \theta}\left(\prod_{i=1}^{N} \sigma_i^z \right)e^{i S_x \theta}.
\end{eqnarray}
For an ideal GHZ state, the expectation value of the parity exhibits high-frequency oscillations,
\begin{eqnarray}
     \expval{\Pi(\theta)}_{\mathrm{GHZ}}=\cos N\theta.
\end{eqnarray}
In Fig.~\ref{fig:2} (a2) and (b2) We plot parity oscillations of the numerically obtained states. The observed oscillatory behavior closely follows the ideal GHZ pattern, confirming the GHZ-like nature of the generated states in our scheme.

For larger systems, full quantum simulations become impractical. Fortunately, the emergence of the GHZ-like state can also be revealed by the time evolution of quantum Fisher information (QFI) in regard to the
generator $S_{x}$, which is proportional to the spin fluctuation along $x$-axis for pure states:
\begin{eqnarray}\label{FQ}
     F_{\mathrm{Q}}^{S_x}\left(\ket{\psi}\right)=4(\Delta S_{x})_{\ket{\psi}}^{2}.
\end{eqnarray}
It quantifies the minimum uncertainty to measure the unknown parameter $\phi$ of a perturbation $e^{i S_{x}\phi}$, namely the quantum Cram\'{e}r-Rao bound
\begin{eqnarray}
     \frac{1}{(\Delta \phi)^2}\leq F_{\mathrm{Q}}^{S_x}.
\end{eqnarray}

The QFI of GHZ-like states should approach the Heisenberg limit $F_{\mathrm{Q}}=N^2$. To numerically study it in large systems, we resort to the discrete truncated Wigner approximation (DTWA) \cite{Schachenmayer2015}. This semi-calssical approach reliably reproduces the time evolution of QFI over the time scales relevant to our study, as demonstrated for small systems in Fig.~\ref{fig:2} (a3) and (b3). In the following sections, we apply DTWA to systematically investigate the performance of our scheme across a wide range of system sizes.

\subsection{The influence of the pulse separation}\label{sec:4a}

\begin{figure}[tbp]
\includegraphics[width=\columnwidth]{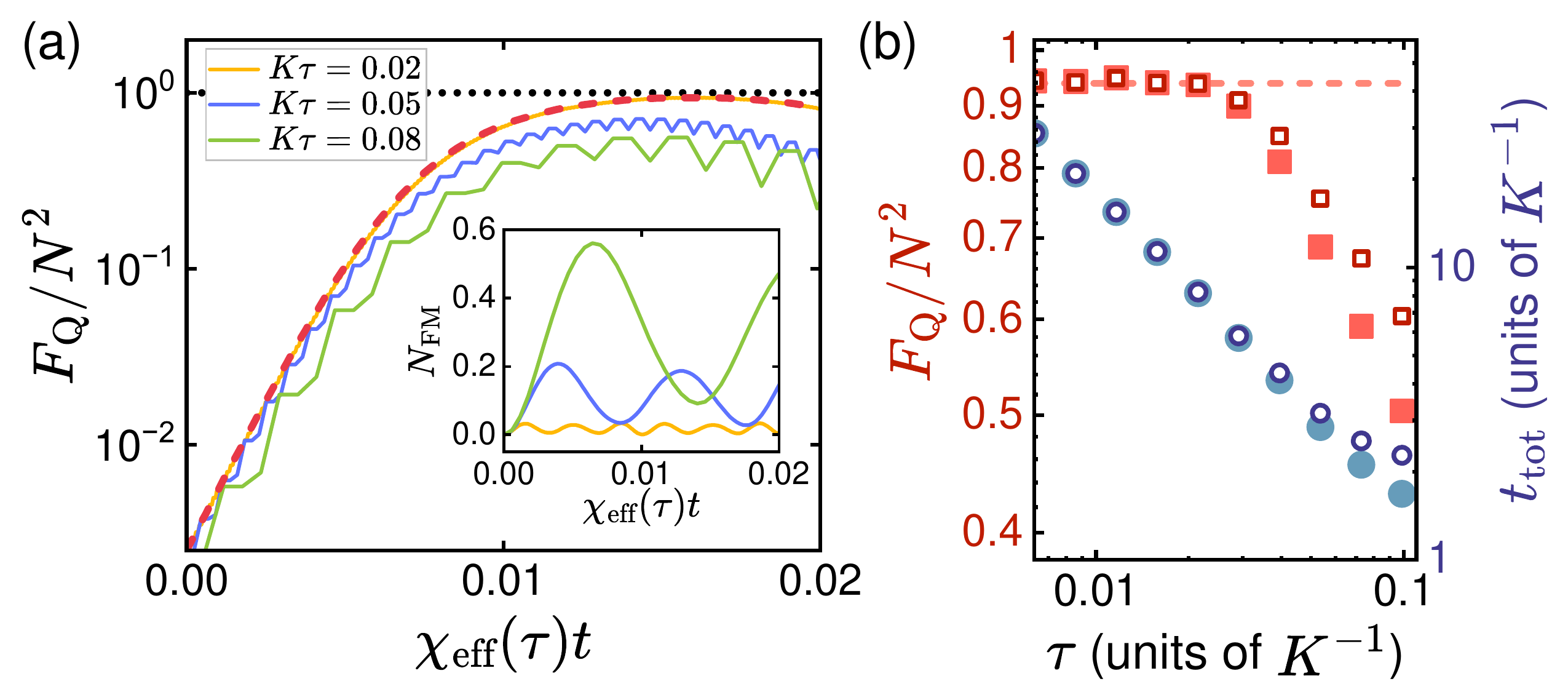}
\caption{The influence of the pulse separation $\tau $ on the
evolution for 2D lattice of $20\times 20$ spins with decaying factor $\alpha =2$. (a) The time evolution of the QFI $F_{\mathrm{Q}}$ for different pulse separations, compared with the result
of the zero-momentum Hamiltonian $H_{\mathrm{ZM}}$ (red dashed curve) and
the Heisenberg limit $F_{\mathrm{Q}}=N^{2}$ (black dotted line). The time
evolution of finite-momentum spin-wave excitation $N_{\mathrm{FM}}$ is shown
in the inset. Here the evolution time has been rescaled with a factor $\chi _{\mathrm{eff}}(\tau )=\lambda NK^{2}\tau /6$. (b) The maximal QFI (red square) and the corresponding total evolution time $t_{\mathrm{tot}}$ (blue circle) versus the pulse separation $\tau $. Open squares (circles) denote the optimal QFI over all generator directions
$F_\mathrm{Q}^\mathrm{opt}$ (and the corresponding evolution time), whereas
solid squares (circles) correspond to the QFI in regard to the generator $S_x$, $F_\mathrm{Q}^{S_x}$ (and the corresponding evolution time).
The red dashed line indicates the maximal $F_{\mathrm{Q}}^{S_x}$ of the
zero-momentum Hamiltonian. The QFI is obtained using the DTWA
(averaged over 1000 trajectories).}
\label{fig:3}
\end{figure}

As discussed in Sec.~\ref{sec:3}, the pulse separation $\tau$ plays a vital role to supress the finite-momentum spin-wave excitations, thereby restoring the dynamics to the collective-spin regime. Here we provide explicit numerical evidence for this effect. In Fig.~\ref{fig:3}(a), considering a 2D lattice of $20\times 20$ spins with interaction decaying factor $\alpha
=2$, we plot the time evolution of QFI and the total finite-momentum
spin-wave excitation
\begin{eqnarray}
     N_{\mathrm{FM}}=\sum_{\bm{q}\neq 0}\expval{b_{\bm{q}}^\dagger b_{\bm{q}}},
\end{eqnarray}
for different pulse separations $\tau $. It clearly
shows that, as $\tau $ decreases, $N_{\mathrm{FM}}$ becomes smaller, and the
evolution of the QFI therefore gets closer to that of the effective
collective-spin cubic XYZ model determined by $H_{\mathrm{ZM}}$ \eqref{H-ZM}, with
the peak value approaching the Heisenberg limit $F_{\mathrm{Q}}=N^{2}$.

The dependence of the maximal QFI and the corresponding total evolution time
$t_{\mathrm{tot}}$ on the pulse separation $\tau $ is provided in Fig.~\ref{fig:3}(b). 
It reveals that when the pulse separation $\tau $ is small
enough, the maximal QFI perfectly matches that of the collective-spin cubic XYZ
model. On the other hand, as the effective interaction strength is
proportional to $\tau $, the total evolution time $t_{\mathrm{tot}}$ becomes
longer for a smaller $\tau $. As a result, there is a trade-off between the
high-fidelity and fast generation of GHZ-like state, and it is crucial to
choose a suitable $\tau $ to ensure the GHZ-like state generation is fast
and efficient.

In addition, we note that the optimal QFI over all generator directions, $F_\mathrm{Q}^\mathrm{opt} = \mathrm{max}_{\bm{n}}\left(F_{\mathrm{Q}}^{S_{\bm{n}}}\right)$, can be determined via the QFI-matrix approach \cite{Reilly2023}. In Fig.~\ref{fig:3}(b) we also show $F_\mathrm{Q}^\mathrm{opt}$ and the corresponding optimal evolution time. In the small-$\tau$ regime, $F_\mathrm{Q}^\mathrm{opt}$ match perfectly with $F_\mathrm{Q}^{S_x}$, indicating that $S_x$ is the optimal generator. For larger $\tau$, $F_\mathrm{Q}^\mathrm{opt}$ becomes slightly larger than $F_\mathrm{Q}^{S_x}$, but their overall behavior remains consistent and leads to no qualitative difference. We therefore conclude that $F_\mathrm{Q}^{S_x}$ provides an accurate and practical characterization of the optimal metrology performance of the states generated in our scheme. In the remainder of this work, we focus exclusively on $F_\mathrm{Q}^{S_x}$.

\subsection{Scaling behavior}\label{sec:4b}

\begin{figure}[tbp]
\includegraphics[width=\columnwidth]{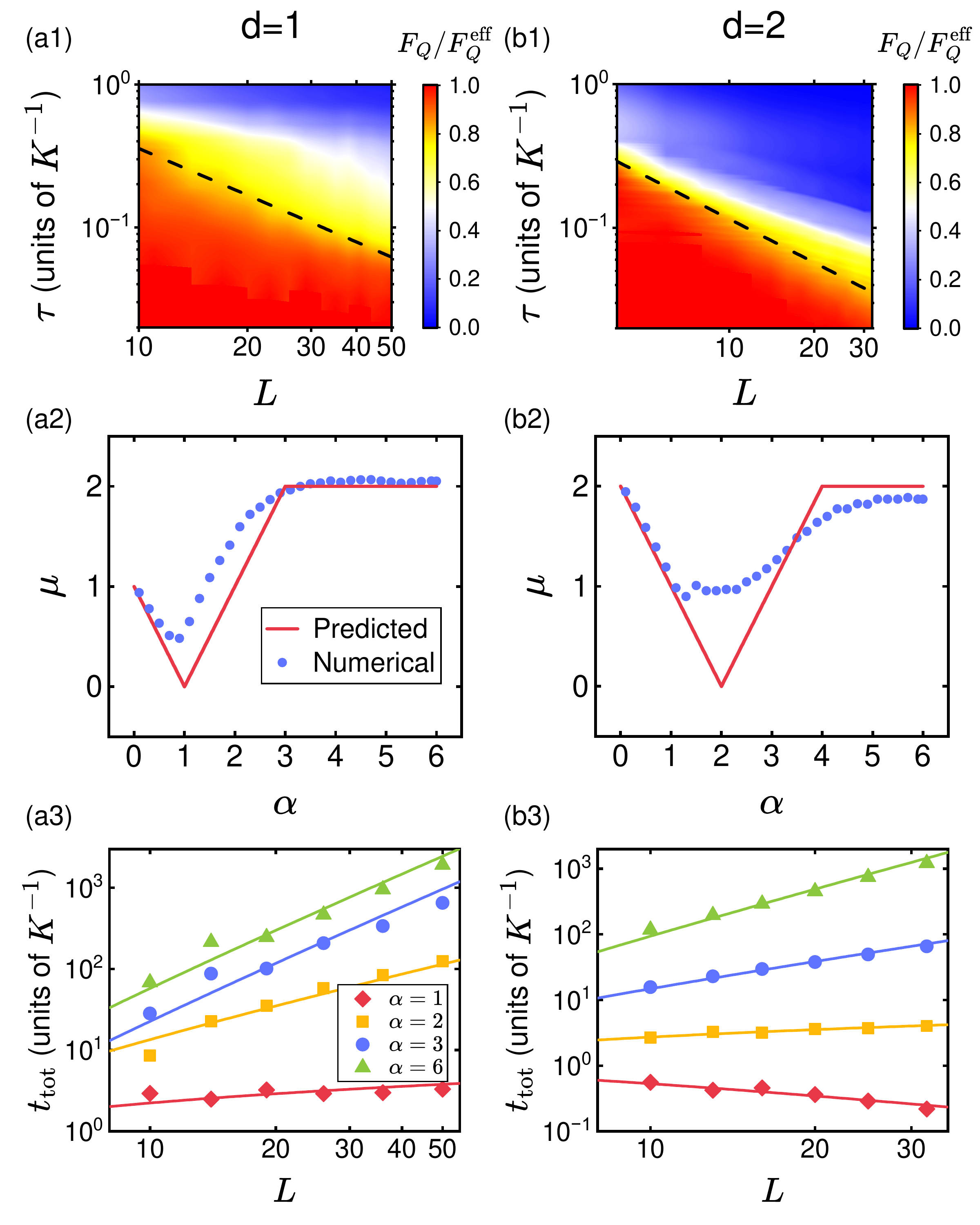}
\caption{The performance of the scheme versus different lattice length $L$,
pulse separation $\protect\tau $ and decaying factor $\protect\alpha $ for
(a) 1D and (b) 2D lattices. (a1, b1) The ratio between the maximal quantum
Fisher information of the scheme and the effective one ($H_{\mathrm{ZM}}$) 
$F_{\mathrm{Q}}/F_{\mathrm{Q}}^{\mathrm{eff}}$ as a function of $L$ and 
$\tau $, calculated at $\protect\alpha =1.5$. The black dashed line
shows the fitting result of $F_{\mathrm{Q}}/F_{\mathrm{Q}}^{\mathrm{eff}}=0.8 $. 
(a2, b2) The power-law exponent $\protect\mu $ of the suitable
pulse separation $\protect\tau _{\mathrm{s}}$ which approximately obeys 
$\tau _{\mathrm{s}}\sim L^{-\protect\mu }$, changing with $\alpha $. 
Both the numerical fitting results (blue circle) and the value predicted
by finite-momentum spin-wave excitations as Eq.~\eqref{mu} (red solid line) are presented. (a3, b3) The total
evolution time as a function of $L$ for different $\protect\alpha $,
compared with the fitted scaling law $t_{\mathrm{tot}}\sim L^{-\protect\nu}\ln L$ 
with $\protect\nu $ given by Eq.~\eqref{nu} (solid lines
with corresponding colors). Numerical results are obtained using the DTWA
(averaged over 1000 trajectories). }
\label{fig:4}
\end{figure}

To indentify a suitable $\tau$ in our scheme, we consider a characteristic pulse separation $\tau_s$ at which the ratio $F_{\mathrm{Q}}/F_{\mathrm{Q}}^{\mathrm{eff}}$ reaches a given threshold, 
where $F_{\mathrm{Q}}^{\mathrm{eff}}$ is the QFI achieved by the effective collective-spin cubic XYZ model \eqref{H-ZM}.
This ratio quantifies the metrological similarity between the generated states under the
synthesized Hamiltonian \eqref{H-eff} and those of $H_{\mathrm{ZM}}$ \eqref{H-ZM}.
Here we use $F_{\mathrm{Q}}/F_{\mathrm{Q}}^{\mathrm{eff}}=0.8$ to study how $\tau_s$ scales with the length of the lattice $L$ for different decay factor $\alpha$ and dimension $d$.
We note that other choices with different $F_{\mathrm{Q}}/F_{\mathrm{Q}}^{\mathrm{eff}}$ give similar results (see Appendix~\ref{app:5}).

From the numerical results we find that the relation between $\tau _{\mathrm{s}}$ and
$L$ approximately follows a power-law form $\tau _{\mathrm{s}}\sim L^{-\mu }$, corresponding to the fitted black dashed line in Fig.~\ref{fig:4}(a1) and
(b1). In Fig.~\ref{fig:4}(a2) and (b2), we plot the numerically fitted $\mu $ as a function of $\alpha$. We can observe an
evident transition at $\alpha =d$, which is often seen as the critical point
that separates short-range and long-range interactions. Remarkably, even for short-range interactions with very large $\alpha $, our
scheme still works and the power-law exponent $\mu $ approaches to 2.

We can gain intuition for the above scaling behavior from the requirement $\tau \ll \mathrm{min}_{\bm{q}}[A_{\bm{q}}/B_{\bm{q}}]$, which implies the scaling law of $\tau_\mathrm{s}$ should follow that of
$\mathrm{min}_{\bm{q}}[A_{\bm{q}}/B_{\bm{q}}]$. Approximating the summations by integrals, we obtain $\tau_\mathrm{s}\sim L^{-\mu }$ with (see
Appendix~\ref{app:3})
\begin{eqnarray}
\mu =\left\{ \begin{aligned} & d-\alpha \quad &\alpha<d, \\ & \alpha-d \quad
&d<\alpha<d+2, \\ & 2 \quad &\alpha\geq d+2. \end{aligned}\right.  \label{mu}
\end{eqnarray}
This matches well with the numerical results as shown in Fig.~\ref{fig:4}(a2) and (b2)
for most parameter ranges. One exception is the transition point near $\alpha =d$, where the dependence changes into a logarithmic one $\tau _{\mathrm{s}}(\alpha =d)\sim (\ln L)^{-2}$.

After determining the pulse separation $\tau _{\mathrm{s}}$, we can obtain the
required evolution time $t_{\mathrm{tot}}$ for generating GHZ-like states, with the results plotted in Fig.~\ref{fig:4}(a3)
and (b3). For its scaling, we can deduce $t_{\mathrm{tot}}\sim L^{-\nu }\ln L$ with
(see Appendix~\ref{app:4})
\begin{eqnarray}
\nu =\left\{ \begin{aligned} & d-\alpha \quad &\alpha<d+2, \\ & -2 \quad
&\alpha\geq d+2. \end{aligned}\right.  \label{nu}
\end{eqnarray}
We can see the fitted scaling law with $\nu$ given by \eqref{nu} indeed captures the behavior of $t_\mathrm{tot}$. Notably, for some parameter ranges, $t_{\mathrm{tot}}$ becomes shorter as $L$ increases, which is especially useful for
generating large-scale GHZ-like states in terms of overcoming dissipations.
From Eq. (\ref{nu}), we can find that this requires $\alpha <(d-1/\ln L)\simeq
d$. Therefore, if the interaction decaying factor is smaller than the system
dimension, fast generation of large-scale GHZ-like states is promising.

\section{Analysis of decoherence}\label{sec:5}

\begin{figure}[tbp]
\includegraphics[width=\columnwidth]{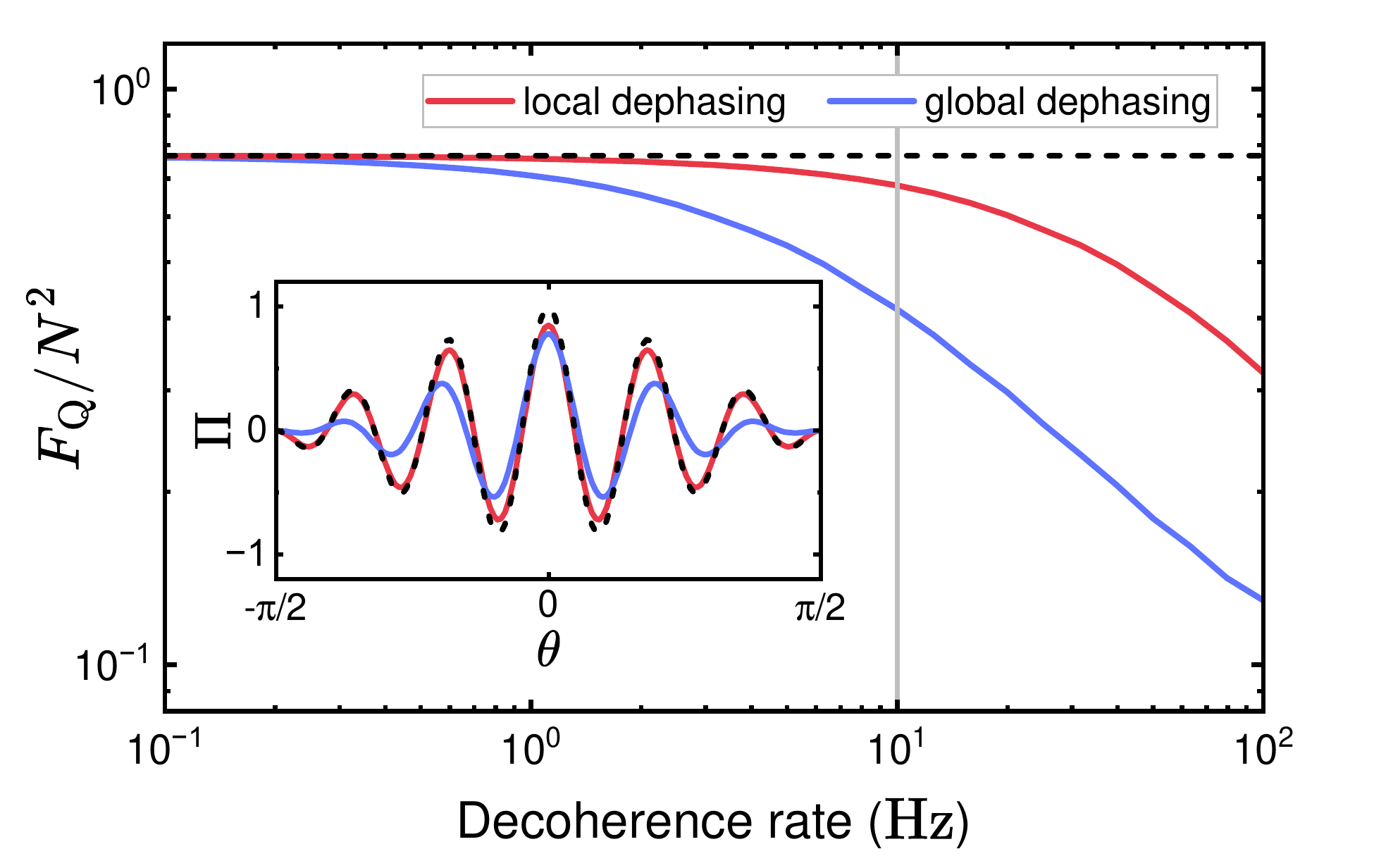}
\caption{The maximal QFI as a function of the decoherence rate for both local ($\gamma$) and global dephasing ($\Gamma$), evaluated under the proposed
experimental parameters ($N=12$ spins in 1D lattice, $K=560\,\mathrm{Hz}$, and $\alpha =1.0$). The pulse separation is fixed at $\tau=0.18\,\text{ms}$. The black dashed line represents the ideal case without decoherence. Inset: expectation value of the parity $\Pi =\prod_{j=1}^{N} \sigma _{j}^{z}$ after applying $e^{i\theta S_{x}}$, calculated at a decoherence rate of $10\text{Hz}$ (gray line). Calculations are based on the exact diagonalization. }
\label{fig:5}
\end{figure}

The preparation of entangled states are often impaded by decoherence in real experiments. In this section, we analyse the potential impact of decoherence by modeling the system dynamics with a Lindblad master equation
\begin{eqnarray}\label{master-eq}
     \frac{d\hat{\rho}}{dt} = -i\left[H, \hat{\rho}\right] + \sum_j \left(L_j \hat{\rho} L_j^\dagger - \frac{1}{2}\left\{L_j^\dagger L_j, \hat{\rho}\right\}\right).
\end{eqnarray}

We focus on two representative types of incoherent noise commonly encountered in experiments: uniformed local dephasing described by the Lindblad operators $L_j=\sqrt{\gamma}s_j^z$, and global dephasing characterized by a collective Lindblad operator $L=\sqrt{\Gamma}S_{z}$. We adopt the experimental parameters of a trapped-ion system as reported in Ref.~\cite{Franke2023}, and vary the decoherence rate of each type of noise respectively.

In open systems, the QFI of a mixed state $\hat{\rho}$ with respect to the generator $S_x$ is evaluated as
\begin{eqnarray}
     F_{\mathrm{Q}}^{S_x}\left(\hat{\rho}\right)=2\sum_{q_{\kappa }+q_{\kappa ^{\prime }}>0}
     \frac{(q_{\kappa }-q_{\kappa ^{\prime }})^{2}}{q_{\kappa }+q_{\kappa
     ^{\prime }}}\abs{\mel{\kappa'}{S_x}{\kappa}}^{2},
\end{eqnarray}
where $\hat{\rho} =\sum_{\kappa }q_{\kappa }\ket{\kappa}\bra{\kappa}$ denotes the
spectral decomposition of the density matrix. We perform full quantum simulation of the master equation \eqref{master-eq} to calculate the QFI, presenting the impact of decoherence in Fig.~\ref{fig:5}.

As shown in Fig.~\ref{fig:5}, local dephasing has a relatively weaker impact as compared to global dephasing. This behavior is expected, since the off-diagonal matrix element $\bra{\downarrow\downarrow...\downarrow}\hat{\rho}\ket{\uparrow\uparrow...\uparrow}$ decays as $e^{-\gamma Nt/2}$ for local dephasing, whereas global dephasing leads to a much faster decay $e^{-\Gamma N^2t/2}$, with an additional factor $N$ aring from the collective coupling to the environment. Furthermore, we find the parity oscillations remain well-resolved despite the presence of moderate decoherence, highlighting the robustness of our scheme against decoherence.

\section{Conclusion}\label{sec:7}
In summary, we explore the scheme for the generation
of large-scale GHZ-like states in lattice spin systems governed by the
power-law Ising model. By utilizing a periodic sequence of global
single-qubit rotational pulses, we can synthesize both three-body
interactions and Heisenberg interactions. The three-body interactions lead
to the creation of GHZ-like states, while the Heisenberg interactions
protect the state evolution in the collective-spin subspace and thus
facilitate the generation of GHZ-like states regardless of the interaction
range. We numerically investigate the scaling behavior of both the suitable pulse separation and the total evolution time for generating the GHZ-like states,
which is intuitively explained by analytically studying finite-momentum spin-wave excitations.
The examination of the performance of our
scheme under different types of decoherence support its robustness. This
work presents a universal and scalable scheme to generate large-scale
GHZ-like states in lattice spin systems, implementable on a variety of
existing platforms with state-of-the-art experimental techniques.

\begin{acknowledgments}
This work is supported by the National Key R\&D Program of China (Grant No.
2023YFA1407600), the National Natural Science Foundation of China (NSFC)
(Grants No. 92576204, No. 12275145, No. 92050110, No. 91736106, No. 11674390, and No.
91836302) and Beijing Key Laboratory of Quantum Artificial Intelligence.
\end{acknowledgments}

\section*{Data availability}
The code that supports the findings of the article are openly available \cite{Code}.

\appendix

\begin{widetext}

\section{Derivation of the effective Hamiltonian}\label{app:1}

Assuming a sufficiently small pulse separation $\tau$, we can expand the evolution operator \eqref{Uoperator} using the Baker-Campbell-Hausdorff (BCH) formula
\begin{eqnarray}
    e^{\epsilon A} e^{\epsilon B} = \exp\left\{\epsilon(A+B) + \frac{1}{2}\epsilon^2[A,B] + \mathcal{O}(\epsilon^2)\right\},
\end{eqnarray}
which gives
\begin{eqnarray}
    U(3\tau) &=& \exp\left\{-i\tau(H_{zz}+H_{xx}) + \frac{1}{2}(-i\tau)^2[H_{zz},H_{xx}] + \mathcal{O}(\tau^2)\right\} \exp\left\{-i\tau H_{yy}\right\} \nonumber \\
    &=& \exp\left\{-i\tau(H_{zz}+H_{xx}+H_{yy}) + \frac{1}{2}(-i\tau)^2 \left([H_{zz},H_{xx}]+[H_{zz},H_{yy}]+[H_{xx},H_{yy}]\right)\right\}.
\end{eqnarray}
The commutators appearing in the 2nd-order term can be calculated as follows:
\begin{eqnarray}
    \left[H_{xx},H_{yy}\right] &=& \sum_{j \neq k}\sum_{m \neq n} K_{jk}K_{mn} [s_j^x s_k^x, s_m^y s_n^y] \nonumber \\
    &=& \sum_{j \neq k}\sum_{m \neq n} K_{jk}K_{mn} \left(s_j^x s_m^y [s_k^x, s_n^y] + s_j^x [s_k^x, s_m^y] s_n^y + [s_j^x, s_m^y] s_n^y s_k^x + s_m^y [s_j^x, s_n^y] s_k^x\right) \nonumber \\
    &=& i \sum_{j \neq k}\sum_{m \neq n} K_{jk}K_{mn} \left(s_j^x s_m^y s_k^z \delta_{k,n} + s_j^x s_k^z s_n^y \delta_{k,m} + s_j^z s_n^y s_k^x \delta_{j,m} + s_m^y s_j^z s_k^x \delta_{j,n}\right) \nonumber \\
    &=& 2i \sum_{[j,k,l]} K_{jl} K_{kl} \left(s_j^x s_k^y s_l^z + s_l^z s_k^y s_j^x\right),
\end{eqnarray}
where the $[j,k,l]$ stands for a combination of mutually unequal $j,k,l$. During the derivation we have used $[s_j^\alpha, s_k^\beta] = i\delta_{jk}\sum_\gamma\epsilon_{\alpha\beta\gamma}s_j^\gamma$ and $\{s_j^\alpha, s_j^\beta\} = \delta_{\alpha\beta}/2$ for spin-$1/2$. Similarly we obtain
\begin{eqnarray}
    \left[H_{zz}, H_{xx}\right] &=& 2i \sum_{[j,k,l]} K_{jk}K_{lk} \left(s_j^x s_k^y s_l^z + s_l^z s_k^y s_j^x\right), \nonumber \\
    \left[H_{zz}, H_{yy}\right] &=& -2i \sum_{[j,k,l]} K_{kj}K _{lj} \left(s_j^x s_k^y s_l^z + s_l^z s_k^y s_j^x\right).
\end{eqnarray}
The evolution operator then reads
\begin{eqnarray}
    U(3\tau) = \exp\left\{-i\tau\left(\sum_{j \neq k}K_{jk}\bm{s}_j\cdot\bm{s}_k + \tau\sum_{[j,k,l]}K_{[j,k,l]}\left(s_j^x s_k^y s_l^z + s_l^z s_k^y s_j^x\right)\right) + \mathcal{O}(\tau^2)\right\},
\end{eqnarray}
where $K_{[j,k,l]} = K_{jl}K_{kl} + K_{jk}K_{lk} - K_{kj}K_{lj}$. Since $U(3\tau) = \exp\{-i3\tau H_\mathrm{eff} + \mathcal{O}(\tau^2)\}$, we obtain the effective Hamiltonian
\begin{eqnarray}
     H_{\mathrm{eff}} &=&\frac{1}{3}\sum_{j\neq k}K_{jk}\bm{s}_{j}\cdot \bm{s}_{k}
     \nonumber \\
     &&+\frac{1}{3}\tau \sum_{\lbrack
     j,k,l\rbrack}K_{[j,k,l]}(s_{j}^{x}s_{k}^{y}s_{l}^{z}+s_{l}^{z}s_{k}^{y}s_{j}^{x}).
\end{eqnarray}

The derivation above requires a sufficiently small pulse separation $\tau$ to ignore higher order terms $\mathcal{O}(\tau^2)$. Since our scheme is focused on the collective part of the model, we can obtain the requirement of $\tau$ by writting down the collective part of the original Hamiltonian \eqref{H-Ising}:
\begin{eqnarray}
    H_{\mathrm{coll}} = \chi_{\mathrm{coll}} S_z^2,
\end{eqnarray}
where the constant terms are neglected and the collective interaction strength reads
\begin{eqnarray}\label{chi-col}
    \chi_{\mathrm{coll}} = \frac{1}{N(N-1)}\sum_{j \neq k}K_{jk}=\frac{K}{N(N-1)}\sum_{j \neq k}\frac{1}{r_{jk}^\alpha}.
\end{eqnarray}
Consider large systems, we can estimate the collective interaction strength by replacing the summation in Eq.~\eqref{chi-col} with an integral. For 1-dimensional lattice the result is
\begin{eqnarray}
    \chi_{\mathrm{coll}}^{d=1} \approx K\frac{2}{L}\int_{1}^{L/2}\mathrm{d}r\frac{1}{r^\alpha} = \left\{
        \begin{aligned}
        & \frac{2K}{(1-\alpha)L}\left[\left(\frac{L}{2}\right)^{1-\alpha} - 1\right] \quad &\alpha\neq1, \\
        & \frac{2K}{L}\ln\left(\frac{L}{2}\right) &\alpha=1.
        \end{aligned}
        \right.
\end{eqnarray}
For 2-dimensional case
\begin{eqnarray}
    \chi_{\mathrm{coll}}^{d=2} \approx K\frac{1}{L^2} 2\pi\int_{1}^{L/2}r\mathrm{d}r\frac{1}{r^\alpha} = \left\{
        \begin{aligned}
        & \frac{2\pi K}{(2-\alpha)L^2}\left[\left(\frac{L}{2}\right)^{2-\alpha}-1\right] &\alpha\neq2,\\
        & \frac{2\pi K}{L^2}\ln\left(\frac{L}{2}\right) &\alpha=2.
        \end{aligned}
        \right.
\end{eqnarray}
It gives the scaling law of the collective interaction strength:
\begin{eqnarray}
    \chi_{\mathrm{coll}} \sim \left\{
        \begin{aligned}
        & K L^{-\alpha} \quad &\alpha<d, \\
        & K L^{-d}\ln L &\alpha=d,\\
        & K L^{-d} &\alpha>d,
        \end{aligned}
        \right.
\end{eqnarray}
where $L$ is the length of the lattice. According to Ref.~\cite{Zhang2024}, the requirement for ignoring higher order terms is given by $\chi_{\mathrm{coll}}\tau N/2\ll 1$. As a result, the critical pulse separation $\tau_{\mathrm{crit}}$ should satisfy
\begin{eqnarray}\label{tau-crit}
    K \tau_{\mathrm{crit}} \sim \left\{
        \begin{aligned}
        & L^{-(d-\alpha)} \quad &\alpha<d, \\
        & \left(\ln L\right)^{-1} &\alpha=d, \\
        & L^{0} &\alpha> d.
        \end{aligned}
        \right.
\end{eqnarray}

\section{Zero-momentum/finite-momentum decomposition}\label{app:2}

In this Appendix, we give the detailed derivation of the zero-momentum/finite-momentum decomposition of the effective Hamiltonian \eqref{ZM-FM}, as well as the analytical expression of the spin-wave excitations \eqref{SW-excit}.

\subsection{Zero-momentum component}

The zero-momentum component of $H_{\mathrm{eff}}$ is its projection in the Dicke manifold $\{\ket{M}\}$:
\begin{eqnarray}
     [H_{\mathrm{eff}}]_{\mathrm{ZM}} = \sum_{M,M'}\mel{M}{H_{\mathrm{eff}}}{M'}\ket{M}\bra{M'}.
\end{eqnarray}
Taking advantage of the permutational symmetry of Dicke states, for $j\neq k$ we have
\begin{eqnarray}
    [\bm{s}_j\cdot\bm{s}_k]_{\mathrm{ZM}} &=& \sum_{M,M'}\mel{M}{\bm{s}_j\cdot\bm{s}_k}{M'}\ket{M}\bra{M'} \nonumber \\
    &=& \frac{1}{N(N-1)} \sum_{M,M'}\mel{M}{\sum_{j \neq k}\bm{s}_j\cdot\bm{s}_k}{M'}\ket{M}\bra{M'},
\end{eqnarray}
and, for mutually distinct $j,k,l$,
\begin{eqnarray}
    [s_j^x s_k^y s_l^z + s_l^z s_k^y s_j^x]_{\mathrm{ZM}} &=& \sum_{M,M'}\mel{M}{s_j^x s_k^y s_l^z + s_l^z s_k^y s_j^x}{M'}\ket{M}\bra{M'} \nonumber \\
    &=& \frac{1}{N(N-1)(N-2)}\sum_{M,M'}\mel{M}{\sum_{[j,k,l]}\left(s_j^x s_k^y s_l^z + s_l^z s_k^y s_j^x\right)}{M'}\ket{M}\bra{M'}.
\end{eqnarray}
Noticed that
\begin{eqnarray}
    \sum_{j \neq k}\bm{s}_j\cdot\bm{s}_k &=& \sum_{j,k}\bm{s}_j\cdot\bm{s}_k - \sum_{j} \bm{s}_j^2 = \bm{S}^2 - \frac{3}{4}N,
\end{eqnarray}
\begin{eqnarray}
     \sum_{[j,k,l]}\left(s_j^x s_k^y s_l^z + s_l^z s_k^y s_j^x\right) &=& \sum_{j,k,l}\left(s_j^x s_k^y s_l^z + s_l^z s_k^y s_j^x\right)- \sum_{j}\left(s_j^x s_j^y s_j^z + s_j^z s_j^y s_j^x\right) \nonumber \\
     & & -\sum_{j \neq k}\left(s_j^x s_k^y s_k^z + s_k^z s_k^y s_j^x\right)-\sum_{j \neq l}\left(s_j^x s_j^y s_l^z + s_l^z s_j^y s_j^x\right) - \sum_{k \neq l}\left(s_l^x s_k^y s_l^z + s_l^z s_k^y s_l^x\right) \nonumber \\
     &=& S_x S_y S_z + S_z S_y S_x,
\end{eqnarray}
after dropping constants we have
\begin{eqnarray}
    H_{\mathrm{ZM}} = [H_{\mathrm{eff}}]_{\mathrm{ZM}} = \frac{1}{3} \lambda K^2\tau \left(J_x J_y J_z + J_z J_y J_x\right),
\end{eqnarray}
where $\bm{J} = \sum_{M,M'}\ket{M}\bra{M}\bm{S}\ket{M'}\bra{M'}$ is a macroscopic spin operator with a fixed spin length $J=N/2$, and

\begin{eqnarray}\label{expression-lambda}
    \lambda &=& \frac{1}{N(N-1)(N-2)}\sum_{[j,k,l]}\left(\frac{1}{r_{jl}^\alpha}\frac{1}{r_{kl}^\alpha} + \frac{1}{r_{jk}^\alpha}\frac{1}{r_{lk}^\alpha} - \frac{1}{r_{kj}^\alpha}\frac{1}{r_{lj}^\alpha}\right) \nonumber \\
    &=& \frac{1}{N(N-1)(N-2)}\sum_{[j,k,l]}\frac{1}{r_{jl}^\alpha}\frac{1}{r_{kl}^\alpha}.
\end{eqnarray}

\subsection{Finite-momentum component}

The finite-momentum component of $H_{\mathrm{eff}}$ is obtained by expressing it in forms of $b_{\bm{q}}$, $b_{\bm{q}}^{\dagger}$ and taking the summation of $\bm{q}\neq0$. For convenience, we explicitly split the effective Hamiltonian as
\begin{eqnarray}
     H_{\mathrm{eff}} = H_0 + H_1,
\end{eqnarray}
including the zeroth-order Heisenberg interaction
\begin{eqnarray}
     H_0 = \frac{1}{3} \sum_{j \neq k}K_{jk}\bm{s}_j\cdot\bm{s}_k,
\end{eqnarray}
and the first-order effective three-body interaction
\begin{eqnarray}
     H_1 = \frac{1}{3}\tau\sum_{[j,k,l]}K_{[j,k,l]}\left(s_j^x s_k^y s_l^z + s_l^z s_k^y s_j^x\right).
\end{eqnarray}
To derive the finite-momentum component, we begin by substituting the HP transformation \eqref{HP-trans} into $H_{\mathrm{eff}}$. Assuming the finite-momentum excitations are weak enough, we keep it up to quadratic terms of the bosonic operators and obtain
\begin{eqnarray}
     H_0 &\simeq& -\frac{1}{12} \sum_{j \neq k}K_{jk} \left[1-2\left(b_j^\dagger b_j + b_k^\dagger b_k\right) + \left(b_j+b_j^\dagger\right)\left(b_k+b_k^\dagger\right) - \left(b_j-b_j^\dagger\right)\left(b_k-b_k^\dagger\right)\right] \nonumber \\
     &=& -\frac{1}{12} \sum_{j \neq k} K_{jk} + \frac{1}{6} \sum_{j \neq k} K_{jk}\left(b_j^\dagger b_j + b_k^\dagger b_k\right) - \frac{1}{6} \sum_{j \neq k} K_{jk}\left(b_j^\dagger b_k + b_k^\dagger b_j\right),
\end{eqnarray}
and
\begin{eqnarray}
     H_1 &\simeq& -i\frac{\tau}{24} \sum_{[j,k,l]} K_{[j,k,l]} \left[\left(b_j+b_j^\dagger\right)\left(b_k-b_k^\dagger\right)+\left(b_k-b_k^\dagger\right)\left(b_j+b_j^\dagger\right)\right] \nonumber \\
     &=& -i\frac{\tau}{12} \sum_{[j,k,l]}K_{[j,k,l]}\left(b_j b_k - b_j^\dagger b_k^\dagger + b_j^\dagger b_k - b_k^\dagger b_j\right).
\end{eqnarray}
The above expression is then rewritten in momentum space as
\begin{eqnarray}\label{H-HP}
    H_0 &\simeq& -\frac{1}{12} \sum_{j \neq k} K_{jk} + \frac{1}{3N} \sum_{j \neq k} K_{jk} \sum_{\bm{q},\bm{q}'}e^{-i\left(\bm{q}-\bm{q}'\right)\cdot \bm{r}_j}b_{\bm{q}}^\dagger b_{\bm{q}'} - \frac{1}{3N} \sum_{j \neq k} K_{jk} \sum_{\bm{q},\bm{q}'}e^{-i\left(\bm{q}\cdot\bm{r}_j-\bm{q}'\cdot\bm{r}_k\right)}b_{\bm{q}}^\dagger b_{\bm{q}'}, \nonumber \\
    H_1 &\simeq& -i\frac{\tau}{12N} \sum_{[j,k,l]} K_{[j,k,l]} \sum_{\bm{q},\bm{q}'}\left[e^{i\left(\bm{q}\cdot\bm{r}_j+\bm{q}'\cdot\bm{r}_k\right)}b_{\bm{q}} b_{\bm{q}'} - e^{-i\left(\bm{q}\cdot\bm{r}_j+\bm{q}'\cdot\bm{r}_k\right)}b_{\bm{q}}^\dagger b_{\bm{q}'}^\dagger + \left(e^{-i\left(\bm{q}\cdot\bm{r}_j-\bm{q}'\cdot\bm{r}_k\right)}-e^{i\left(\bm{q}\cdot\bm{r}_j-\bm{q}'\cdot\bm{r}_k\right)}\right)b_{\bm{q}}^\dagger b_{\bm{q}'}\right].\nonumber \\
\end{eqnarray}
For periodic lattices we may simplify the expression as
\begin{eqnarray}
     \frac{1}{N}\sum_{j \neq k}K_{jk}\sum_{\bm{q},\bm{q}'}e^{-i\left(\bm{q}-\bm{q}'\right)\cdot \bm{r}_j}b_{\bm{q}}^\dagger b_{\bm{q}'} = \sum_{\bm{q}} K_0 b_{\bm{q}}^\dagger b_{\bm{q}},
\end{eqnarray}
\begin{eqnarray}
     \frac{1}{N}\sum_{j \neq k}K_{jk}\sum_{\bm{q},\bm{q}'}e^{-i\left(\bm{q}\cdot \bm{r}_j-\bm{q}'\cdot \bm{r}_k\right)}b_{\bm{q}}^\dagger b_{\bm{q}'} = \sum_{\bm{q}}K_{\bm{q}} b_{\bm{q}}^\dagger b_{\bm{q}},
\end{eqnarray}
where
\begin{eqnarray}
     K_{\bm{q}} = K\sum_{\bm{r}\neq 0}\frac{1}{r^\alpha}e^{-i\bm{q}\cdot\bm{r}}.
\end{eqnarray}
Also we have
\begin{eqnarray}
     \frac{1}{N}\sum_{[j,k,l]} K_{[j,k,l]} \sum_{\bm{q},\bm{q}'}e^{i\left(\bm{q}\cdot\bm{r}_j+\bm{q}'\cdot\bm{r}_k\right)}b_{\bm{q}} b_{\bm{q}'} = \sum_{\bm{q}}\left(K_{\bm{q}}^2-T_0^2\right) b_{\bm{q}}b_{-\bm{q}},
\end{eqnarray}
\begin{eqnarray}
     \frac{1}{N}\sum_{[j,k,l]} K_{[j,k,l]} \sum_{\bm{q},\bm{q}'}e^{-i\left(\bm{q}\cdot\bm{r}_j+\bm{q}'\cdot\bm{r}_k\right)}b_{\bm{q}}^\dagger b_{\bm{q}'}^\dagger = \sum_{\bm{q}}\left(K_{\bm{q}}^2-T_0^2\right) b_{\bm{q}}^{\dagger}b_{-\bm{q}}^{\dagger},
\end{eqnarray}
\begin{eqnarray}
     \frac{1}{N}\sum_{[j,k,l]} K_{[j,k,l]} \sum_{\bm{q},\bm{q}'}e^{-i\left(\bm{q}\cdot\bm{r}_j-\bm{q}'\cdot\bm{r}_k\right)}b_{\bm{q}}^\dagger b_{\bm{q}'} = \sum_{\bm{q}}\left(K_{\bm{q}}^2-T_0^2\right) b_{\bm{q}}^{\dagger}b_{\bm{q}},
\end{eqnarray}
\begin{eqnarray}
     \frac{1}{N}\sum_{[j,k,l]} K_{[j,k,l]} \sum_{\bm{q},\bm{q}'}e^{i\left(\bm{q}\cdot\bm{r}_j-\bm{q}'\cdot\bm{r}_k\right)}b_{\bm{q}}^\dagger b_{\bm{q}'} = \sum_{\bm{q}}\left(K_{\bm{q}}^2-T_0^2\right) b_{\bm{q}}^{\dagger}b_{\bm{q}},
\end{eqnarray}
where
\begin{eqnarray}
     T_0^2 = K^2\sum_{\bm{r}\neq 0}\frac{1}{r^{2\alpha}}.
\end{eqnarray}
Taking the summation over the $\bm{q}\neq0$ sector, we arrive at the expression of the finite-momentum component
\begin{eqnarray}
     H_{\mathrm{SW}} &=&-\frac{N}{12}K_{0}+\frac{1}{3}\sum_{\bm{q}\neq 0}\left(
     K_{0}-K_{\bm{q}}\right) b_{\bm{q}}^{\dagger }b_{\bm{q}}  \nonumber \\
     &&-i\frac{\tau }{12}\sum_{\bm{q}\neq 0}\left( K_{\bm{q}}^{2}-T_{0}^{2}\right) \left( b_{\bm{q}}b_{-\bm{q}}-b_{\bm{q}}^{\dagger }b_{-\bm{q}}^{\dagger }\right),\nonumber \\
\end{eqnarray}

\subsection{Bogoliubov diagonalization of the spin-wave Hamiltonian}

The spin-wave Hamiltonian \eqref{H-SW} can be excatly diagonalized via Bogoliubov transformation. We first write it in the quadratic form
\begin{eqnarray}
    H_{\mathrm{SW}} = \frac{1}{2}\sum_{\bm{q}\neq 0}
    \begin{pmatrix}
        \beta_{\bm{q}}^\dagger \\
        \beta_{-\bm{q}}
    \end{pmatrix}^T
    \begin{pmatrix}
        A_{\bm{q}} & \tau B_{\bm{q}} \\
        \tau B_{\bm{q}} & A_{\bm{q}} \\
    \end{pmatrix}
    \begin{pmatrix}
        \beta_{\bm{q}} \\
        \beta_{-\bm{q}}^\dagger
    \end{pmatrix},
\end{eqnarray}
where we have neglected the constant term and defined
\begin{eqnarray}
     A_{\bm{q}} &=& \frac{1}{3}\left(K_0 - K_{\bm{q}}\right), \qquad B_{\bm{q}} = \frac{1}{6}\left(K_{\bm{q}}^2-T_0^2\right), \nonumber\\
     \beta_{\bm{q}} &=& e^{-i\frac{\pi}{4}}b_{\bm{q}}.
\end{eqnarray}
We introduce the Bogoliubov transformation $\beta_{\bm{q}} = (u_{\bm{q}}a_{\bm{q}} - v_{\bm{q}}a_{-\bm{q}}^\dagger)$ with
\begin{eqnarray}
     u_{\bm{q}} &=& \sqrt{\frac{1}{2}\left(\frac{A_{\bm{q}}}{\epsilon_{\bm{q}}}+1\right)},\nonumber\\ 
     v_{\bm{q}} &=& \mathrm{sgn}\left(B_{\bm{q}}\right)\sqrt{\frac{1}{2}\left(\frac{A_{\bm{q}}}{\epsilon_{\bm{q}}}-1\right)},
\end{eqnarray}
where the excitation energy $\epsilon_{\bm{q}} = \sqrt{A_{\bm{q}}^2-\tau^2 B_{\bm{q}}^2}$. The spin-wave Hamiltonian is then written as
\begin{eqnarray}
     H_{\mathrm{SW}} = \sum_{\bm{q}\neq0}\epsilon_{\bm{q}} \left(a_{\bm{q}}^\dagger a_{\bm{q}} + \frac{1}{2}\right).
\end{eqnarray}
As a result, we can explicitly write down the time-dependence of $a_{\bm{q}}$ as
\begin{eqnarray}
     a_{\bm{q}}(t) = a_{\bm{q}}(0) e^{-i\epsilon_{\bm{q}}t}.
\end{eqnarray}
Since $\epsilon_{\bm{q}}=\epsilon_{-\bm{q}}$, we have
\begin{eqnarray}\label{b-excit}
     \expval{b_{\bm{q}}^\dagger b_{\bm{q}}}_t &=& \expval{\beta_{\bm{q}}^\dagger \beta_{\bm{q}}}_t \nonumber \\
     &=& u_{\bm{q}}^2 \expval{a_{\bm{q}}^\dagger a_{\bm{q}}}_0 + v_{\bm{q}}^2 \expval{a_{-\bm{q}} a_{-\bm{q}}^\dagger}_0 \nonumber\\
     && - u_{\bm{q}}v_{\bm{q}} \left(\expval{a_{\bm{q}}^\dagger a_{-\bm{q}}^\dagger}_0e^{2i\epsilon_{\bm{q}}t} + \expval{a_{-\bm{q}} a_{\bm{q}}}_0e^{-2i\epsilon_{\bm{q}}t}\right).\nonumber\\
\end{eqnarray}
Using the inverse transformation $a_{\bm{q}} = u_{\bm{q}}\beta_{\bm{q}}+v_{\bm{q}}\beta_{-\bm{q}}^\dagger$ we obtain
\begin{eqnarray}
     \expval{a_{\bm{q}}^\dagger a_{\bm{q}}}_0 &=& u_{\bm{q}}^2 \expval{\beta_{\bm{q}}^\dagger\beta_{\bm{q}}}_0 + v_{\bm{q}}^2 \expval{\beta_{-\bm{q}}\beta_{-\bm{q}}^\dagger}_0 \nonumber\\
     && + u_{\bm{q}}v_{\bm{q}} \left(\expval{\beta_{\bm{q}}^\dagger\beta_{-\bm{q}}^\dagger}_0 + \expval{\beta_{-\bm{q}}\beta_{\bm{q}}}_0\right)\nonumber. \\
\end{eqnarray}
Since the initial coherent spin state $\ket{\uparrow}^{\otimes N}$ is the vacuum state of all $b_{\bm{q}}$, the above expression reduces to
\begin{eqnarray}
     \expval{a_{\bm{q}}^\dagger a_{\bm{q}}}_0 = v_{\bm{q}}^2.
\end{eqnarray}
Similarly we have
\begin{eqnarray}
     \expval{a_{-\bm{q}} a_{-\bm{q}}^\dagger}_0 = u_{\bm{q}}^2, \quad \expval{a_{\bm{q}}^\dagger a_{-\bm{q}}^\dagger}_0 = \expval{a_{-\bm{q}} a_{\bm{q}}}_0 = u_{\bm{q}}v_{\bm{q}}. \nonumber\\
\end{eqnarray}
Substituting them into Eq.~\eqref{b-excit}, we arrive at the explicit expression of finite-momentum spin-wave excitations:
\begin{eqnarray}
     \expval{b_{\bm{q}}^\dagger b_{\bm{q}}}_t &=& 2 u_{\bm{q}}^2 v_{\bm{q}}^2 \left[1-\cos\left(2\epsilon_{\bm{q}}t\right)\right]\nonumber\\
     &=& \frac{\tau^2}{2}\frac{B_{\bm{q}}^2}{A_{\bm{q}}^2 - \tau^2 B_{\bm{q}}^2}\left[1-\cos\left(2\epsilon_{\bm{q}}t\right)\right].
\end{eqnarray}

\section{The scaling law of the suitable pulse separation}\label{app:3}

The requirement for ignoring finite-momentum excitations is $\tau\ll \mathrm{min}_{\bm{q}}[A_{\bm{q}}/B_{\bm{q}}]$, suggesting that the suitable pulse separation $\tau_{\mathrm{s}}$ has the same scaling law with $\mathrm{min}_{\bm{q}}[A_{\bm{q}}/B_{\bm{q}}]$. It can be figured out that the mode with the minimum momentum $\bm{q_1}=(2\pi/L,0,0,...,0)$ has the largest excitation. In the following, we study the scaling law of $\mathrm{min}_{\bm{q}}[A_{\bm{q}}/B_{\bm{q}}] = A_{\bm{q_1}}/B_{\bm{q_1}}$ with the help of continuous approximation
\begin{eqnarray}\label{Kq-int}
    K_0 &=& K\sum_{\bm{r}\neq 0}\frac{1}{r^\alpha} \approx K \varepsilon^{\alpha-d}\int_{\bm{x}\in Z_1^d\setminus Z_\varepsilon^d} \mathrm{d}^d x \frac{1}{x^\alpha}, \nonumber\\
    K_{\bm{q_1}} &=& K\sum_{\bm{r}\neq 0}\frac{1}{r^\alpha}e^{-i\bm{q_1}\cdot\bm{r}} \approx K \varepsilon^{\alpha-d}\int_{\bm{x}\in Z_1^d\setminus Z_\varepsilon^d} \mathrm{d}^d x \frac{1}{x^\alpha}e^{-i\pi x_1},\nonumber\\
    T_0^2 &=& K^2 \sum_{\bm{r}\neq 0}\frac{1}{r^{2\alpha}} \approx K^2 \varepsilon^{2\alpha-d}\int_{\bm{x}\in Z_1^d\setminus Z_\varepsilon^d} \mathrm{d}^d x \frac{1}{x^{2\alpha}},
\end{eqnarray}
where $\varepsilon = 2/L$ and $Z_{y} = [-y,y]$.

\subsection{1-dimensional case}

For $d=1$ we have
\begin{eqnarray}
    K_0 &\approx& K \varepsilon^{\alpha-1}2\int_{\epsilon}^1 \mathrm{d}x \frac{1}{x^\alpha} = K\varepsilon^{\alpha-1}\frac{2}{1-\alpha}\left(1-\varepsilon^{1-\alpha}\right), \nonumber\\
    K_{\bm{q_1}} &\approx& K \varepsilon^{\alpha-1}2\int_{\epsilon}^1 \mathrm{d}x \frac{1}{x^\alpha}\cos\left(\pi x\right) \nonumber \\
    &=& K\varepsilon^{\alpha-1}\frac{-i}{\pi}\left\{\left(-i\pi\right)^\alpha\left[\Gamma\left(1-\alpha,-i\pi\right)-\Gamma\left(1-\alpha,-i\pi\varepsilon\right)\right]-\left(i\pi\right)^\alpha\left[\Gamma\left(1-\alpha,i\pi\right)-\Gamma\left(1-\alpha,i\pi\varepsilon\right)\right]\right\}, \nonumber \\
    T_0^2 &\approx& K^2 \varepsilon^{2\alpha-1}2\int_{\varepsilon}^{1} \mathrm{d}x \frac{1}{x^{2\alpha}} = K^2 \varepsilon^{2\alpha-1}\frac{2}{1-2\alpha}\left(1-\varepsilon^{1-2\alpha}\right),
\end{eqnarray}
where the incomplete Gamma function is defined as
\begin{eqnarray}
    \Gamma\left(a,z\right) = \int_{z}^{\infty} t^{a-1}e^{-t}\mathrm{d}t.
\end{eqnarray}
Utilizing the expansion
\begin{eqnarray}
    \Gamma\left(1-\alpha, -i\pi\epsilon\right) = \Gamma\left(1-\alpha\right) + \frac{\left(-i\pi\varepsilon\right)^{1-\alpha}}{\alpha-1}-\frac{\left(-i\pi\varepsilon\right)^{2-\alpha}}{\alpha-2}+ \frac{\left(-i\pi\varepsilon\right)^{3-\alpha}}{2\left(\alpha-3\right)} + \mathcal{O}\left(\varepsilon^{4-\alpha}\right),
\end{eqnarray}
we can obtain the asymptotic behaviour of $A_{\bm{q_1}}/B_{\bm{q_1}}$ as
\begin{eqnarray}\label{scal-1d}
    \frac{A_{\bm{q_1}}}{B_{\bm{q_1}}} &=& 2\frac{K_0 - K_{\bm{q_1}}}{K_{\bm{q_1}}^2 - T_0^2} \nonumber \\
    &=& \frac{2}{K}\varepsilon^{1-\alpha} \frac{\frac{2}{1-\alpha}+E_\alpha\left(i\pi\right)+E_\alpha\left(-i\pi\right)-2\cos\left(\frac{\alpha-1}{2}\pi\right)\pi^{\alpha-1}\Gamma\left(1-\alpha\right) + \frac{\pi^2}{\alpha-3}\varepsilon^{3-\alpha} + \mathcal{O}\left(\varepsilon^{4-\alpha}\right)}{\left[E_\alpha\left(i\pi\right)+E_\alpha\left(-i\pi\right)-2\cos\left(\frac{\alpha-1}{2}\pi\right)\pi^{\alpha-1}\Gamma\left(1-\alpha\right)+\frac{2}{1-\alpha}\varepsilon^{1-\alpha}+\frac{\pi^2}{\alpha-3}\varepsilon^{3-\alpha}+\mathcal{O}\left(\varepsilon^{4-\alpha}\right)\right]^2 - \frac{2}{1-2\alpha}\varepsilon\left(1-\varepsilon^{1-2\alpha}\right)}, \nonumber \\
\end{eqnarray}
where we use the exponential integral function
\begin{eqnarray}
    E_n(z)=\int_{1}^{\infty}\frac{e^{-zt}}{t^n}\mathrm{d}t.
\end{eqnarray}

For $\alpha<1$, the lowest-order term in Eq.~\eqref{scal-1d} is
\begin{eqnarray}
    \frac{A_{\bm{q_1}}}{B_{\bm{q_1}}} \sim \frac{2}{K}\frac{\frac{2}{1-\alpha}+E_\alpha\left(i\pi\right)+E_\alpha\left(-i\pi\right)-2\cos\left(\frac{\alpha-1}{2}\pi\right)\pi^{\alpha-1}\Gamma\left(1-\alpha\right)}{\left[E_\alpha\left(i\pi\right)+E_\alpha\left(-i\pi\right)-2\cos\left(\frac{\alpha-1}{2}\pi\right)\pi^{\alpha-1}\Gamma\left(1-\alpha\right)\right]^2} \varepsilon^{1-\alpha} \sim K^{-1} L^{\alpha-1},
\end{eqnarray}
for $1<\alpha<3$ it is
\begin{eqnarray}
    \frac{A_{\bm{q_1}}}{B_{\bm{q_1}}} \sim \frac{2}{K}\frac{\frac{2}{1-\alpha}+E_\alpha\left(i\pi\right)+E_\alpha\left(-i\pi\right)-2\cos\left(\frac{\alpha-1}{2}\pi\right)\pi^{\alpha-1}\Gamma\left(1-\alpha\right)}{\frac{4}{(\alpha-1)^2}-\frac{2}{2\alpha-1}} \varepsilon^{\alpha-1} \sim K^{-1} L^{1-\alpha},
\end{eqnarray}
and for $\alpha>3$
\begin{eqnarray}
    \frac{A_{\bm{q_1}}}{B_{\bm{q_1}}} \sim \frac{2}{K}\frac{\frac{\pi^2}{\alpha-3}}{\frac{4}{(\alpha-1)^2}-\frac{2}{2\alpha-1}} \varepsilon^{2} \sim K^{-1} L^{-2}.
\end{eqnarray}
We conclude the scaling law in 1-dimensional case as
\begin{equation}\label{mu-1d}
     K\tau_{\mathrm{s}}^{d=1}\sim\left\{ \begin{aligned} & L^{-(1-\alpha)} \quad &\alpha<1, \\ & L^{-(\alpha-1)} \quad
     &1<\alpha<3, \\ & L^{-2} \quad &\alpha\geq 3. \end{aligned}\right.
\end{equation}

Note that at the transition point $\alpha=d=1$, the result is special:
\begin{eqnarray}
    K_0 &\approx& 2K \int_{\epsilon}^1 \mathrm{d}x \frac{1}{x} = -2K\ln\varepsilon, \nonumber\\
    K_{\bm{q_1}} &\approx& 2K \int_{\epsilon}^1 \mathrm{d}x \frac{1}{x}\cos\left(\pi x\right) = 2K\left[\mathrm{Ci}(\pi)-\mathrm{Ci}(\pi\varepsilon)\right], \nonumber \\
    T_0^2 &\approx& 2K^2 \varepsilon\int_{\varepsilon}^{1} \mathrm{d}x \frac{1}{x^{2}} = 2K^2 (1-\varepsilon),
\end{eqnarray}
where the cosine integral function is defined as
\begin{eqnarray}
    \mathrm{Ci}(z) = -\int_{z}^{\infty}\frac{\cos(t)}{t}\mathrm{d}t,
\end{eqnarray}
which has the expansion
\begin{eqnarray}
    \mathrm{Ci}(\pi\varepsilon) = \ln(\pi\varepsilon) + \gamma_{\mathrm{Euler}} + \mathrm{O}(\varepsilon^2),
\end{eqnarray}
where $\gamma_{\mathrm{Euler}}\approx0.577$ is the Euler constant. The lowest-order term is now given by
\begin{eqnarray}
    \frac{A_{\bm{q_1}}}{B_{\bm{q_1}}} \sim \frac{1}{K} \frac{\ln\pi+\gamma_{\mathrm{Euler}}-\mathrm{Ci}(\pi)}{(\ln L)^2} \sim K^{-1} (\ln L)^{-2}.
\end{eqnarray}

\subsection{2-dimensional case}

In the case of $d=2$ we have
\begin{eqnarray}
     K_0 &\approx& K \varepsilon^{\alpha-2}2\pi\int_{\epsilon}^1 r\mathrm{d}r \frac{1}{r^\alpha} = K\frac{2\pi}{2-\alpha}\varepsilon^{\alpha-2}\left(1-\varepsilon^{2-\alpha}\right), \nonumber\\
     K_{\bm{q_1}} &\approx& K \varepsilon^{\alpha-2}\int_{\epsilon}^1 r\mathrm{d}r \frac{1}{r^\alpha}\int_{0}^{2\pi}d\theta e^{-i\pi r\cos\theta} =K\varepsilon^{\alpha-2}2\pi\int_{\varepsilon}^{1}r \mathrm{d}r \frac{1}{r^\alpha}J_0(\pi r) \nonumber \\
     &=& K\frac{2\pi}{2-\alpha}\varepsilon^{\alpha-2}\left[{}_{1}F_2\left(1-\frac{\alpha}{2};1,2-\frac{\alpha}{2};-\frac{\pi^2}{4}\right) - \varepsilon^{2-\alpha}{}_{1}F_2\left(1-\frac{\alpha}{2};1,2-\frac{\alpha}{2};-\frac{\pi^2}{4}\varepsilon^2\right)\right], \nonumber \\
     T_0^2 &\approx& K^2 \varepsilon^{2\alpha-2}2\pi\int_{\varepsilon}^{1} r\mathrm{d}r \frac{1}{r^{2\alpha}} = K^2\frac{\pi}{1-\alpha}\varepsilon^{2\alpha-2}\left(1-\varepsilon^{2-2\alpha}\right),
\end{eqnarray}
where we have approximately replaced the region $\bm{x}\in Z_1^2\setminus Z_\varepsilon^2$ with $\varepsilon<\abs{\bm{x}}<1$ to simplify the integral, $J_n(z)$ is the nth-order Bessel function of the first kind, and ${}_{p}F_{q}\left(\left\{a\right\};\left\{b\right\};z\right)$ is the generalized hypergeometric function
\begin{eqnarray}
    {}_{p}F_{q}\left(\left\{a\right\};\left\{b\right\};z\right) = \sum_{k=0}^{\infty}\frac{(a_1)_k(a_2)_k...(a_p)_k}{(b_1)_k(b_2)_k...(b_q)_k}\frac{z^k}{k!},
\end{eqnarray}
which is defined with the Pochhammer symbol
\begin{eqnarray}
    (a)_n = \left\{
		\begin{aligned}
		& 1 \quad &n=0, \\
		& a(a+1)(a+2)...(a+n-1) \quad &n>0. \\
		\end{aligned}
		\right.
\end{eqnarray}
We can expand the expression using
\begin{eqnarray}
    {}_{1}F_{2}\left(1-\frac{\alpha}{2};1,2-\frac{\alpha}{2};-\frac{1}{4}\pi^2\varepsilon^2\right) = 1-\frac{\alpha-2}{\alpha-4}\frac{\pi^2}{4}\varepsilon^2 + \mathcal{O}\left(\varepsilon^4\right),
\end{eqnarray}
so that
\begin{eqnarray}\label{scal-2d}
    \frac{A_{\bm{q_1}}}{B_{\bm{q_1}}} &=& 2\frac{K_0 - K_{\bm{q_1}}}{K_{\bm{q_1}}^2 - T_0^2} \nonumber \\
    &=& \frac{2}{\pi K}\varepsilon^{2-\alpha} \frac{\frac{2}{2-\alpha}\left[1-{}_{1}F_{2}\left(1-\frac{\alpha}{2};1,2-\frac{\alpha}{2};-\frac{\pi^2}{4}\right)\right]+\frac{2}{\alpha-4}\frac{\pi^2}{4}\varepsilon^{4-\alpha}+\mathcal{O}\left(\varepsilon^{6-\alpha}\right)}{\frac{4}{(2-\alpha)^2}\left[{}_{1}F_{2}\left(1-\frac{\alpha}{2};1,2-\frac{\alpha}{2};-\frac{\pi^2}{4}\right)-\varepsilon^{2-\alpha}+\frac{\alpha-2}{\alpha-4}\frac{\pi^2}{4}\varepsilon^{4-\alpha}+\mathcal{O}\left(\varepsilon^{6-\alpha}\right)\right]^2 - \frac{1}{(1-\alpha)\pi}\varepsilon^2\left(1-\varepsilon^{2-2\alpha}\right)}. \nonumber \\
\end{eqnarray}

For $\alpha<2$, the lowest order term is given by
\begin{eqnarray}
    \frac{A_{\bm{q_1}}}{B_{\bm{q_1}}} &\sim& \frac{2-\alpha}{2\pi K}\frac{1-{}_{1}F_{2}\left(1-\frac{\alpha}{2};1,2-\frac{\alpha}{2};-\frac{\pi^2}{4}\right)}{\left[{}_{1}F_{2}\left(1-\frac{\alpha}{2};1,2-\frac{\alpha}{2};-\frac{\pi^2}{4}\right)\right]^2}\varepsilon^{2-\alpha}\nonumber\\
    &\sim& K^{-1} L^{\alpha-2},
\end{eqnarray}
for $2<\alpha<4$ it is
\begin{eqnarray}
    \frac{A_{\bm{q_1}}}{B_{\bm{q_1}}} &\sim& \frac{2}{(2-\alpha)\pi K}\frac{1-{}_{1}F_{2}\left(1-\frac{\alpha}{2};1,2-\frac{\alpha}{2};-\frac{\pi^2}{4}\right)}{\frac{4}{(\alpha-2)^2}-\frac{1}{(\alpha-1)\pi}}\varepsilon^{\alpha-2}\nonumber\\
    &\sim& K^{-1} L^{2-\alpha},
\end{eqnarray}
and for $\alpha>4$
\begin{eqnarray}
    \frac{A_{\bm{q_1}}}{B_{\bm{q_1}}} \sim \frac{2}{(\alpha-4)\pi K}\frac{\frac{\pi^2}{4}}{\frac{4}{(\alpha-2)^2}-\frac{1}{(\alpha-1)\pi}} \varepsilon^{2} \sim K^{-1} L^{-2}\nonumber\\.
\end{eqnarray}
The scaling law for 2-dimensional case is
\begin{equation}\label{mu-2d}
     K\tau_{\mathrm{s}}^{d=2}\sim\left\{ \begin{aligned} & L^{-(2-\alpha)} \quad &\alpha<2, \\ & L^{-(\alpha-2)} \quad
     &2<\alpha<4, \\ & L^{-2} \quad &\alpha\geq 4. \end{aligned}\right.
\end{equation}

At the transition point $\alpha=d=2$, the result becomes
\begin{eqnarray}
    K_0 &\approx& 2\pi K \int_{\epsilon}^1 \mathrm{d}x \frac{1}{r} = -2\pi K\ln\varepsilon, \nonumber\\
    K_{\bm{q_1}} &\approx& 2\pi K \int_{\epsilon}^1 \mathrm{d}r \frac{1}{r}J_0(\pi r) = -\pi K\left[G_{1,3}^{2,0}\left(\frac{\pi^2}{4}|_{0,0,0}^1\right)-G_{1,3}^{2,0}\left(\frac{\pi^2\varepsilon^2}{4}|_{0,0,0}^1\right)\right], \nonumber \\
    T_0^2 &\approx& 2\pi K^2 \varepsilon^2\int_{\varepsilon}^{1} \mathrm{d}r \frac{1}{r^{3}} = \pi K^2 (1-\varepsilon^2),
\end{eqnarray}
where $G_{p,q}^{m,n}\left(z|_{b_1,...,b_q}^{a_1,...,a_p}\right)$ is the Meijer G-function
\begin{eqnarray}
    G_{p,q}^{m,n}\left(z|_{b_1,...,b_q}^{a_1,...,a_p}\right) = \frac{1}{2\pi i}\int\frac{\Gamma(1-a_1-s)...\Gamma(1-a_n-s)\Gamma(b_1+s)...\Gamma(b_m+s)}{\Gamma(a_{n+1}+s)...\Gamma(a_p+s)\Gamma(1-b_{m+1}-s)...\Gamma(1-b_p-s)}z^{-s}\mathrm{d}s,
\end{eqnarray}
which can be expanded as
\begin{eqnarray}
    G_{1,3}^{2,0}\left(\frac{\pi^2\varepsilon^2}{4}|_{0,0,0}^1\right) = -2\ln\left(\frac{\pi}{2}\varepsilon\right) - 2\gamma_{\mathrm{Euler}} + \mathcal{O}\left(\varepsilon^2\right).
\end{eqnarray}
Thus we obtain
\begin{eqnarray}
    \frac{A_{\bm{q_1}}}{B_{\bm{q_1}}} \sim \frac{1}{\pi K} \frac{\ln\frac{\pi}{2}+\gamma_{\mathrm{Euler}}+\frac{1}{2}G_{1,3}^{2,0}\left(\frac{\pi^2}{4}|_{0,0,0}^1\right)}{(\ln L)^2} \sim K^{-1} (\ln L)^{-2}.
\end{eqnarray}

\subsection{The uniform expression}
It is easy to find the scaling law of $\tau_{\mathrm{s}}$--for both $d=1$ and $d=2$--can be expressed in a uniform way

\begin{eqnarray}
	K\tau_{\mathrm{s}} \sim \left\{
		\begin{aligned}
		& L^{-(d-\alpha)} \quad &\alpha<d, \\
		& L^{-(\alpha-d)} \quad &d<\alpha<d+2, \\
		& L^{-2} \quad &\alpha\geq d+2.
		\end{aligned}
		\right.
\end{eqnarray}
At the transition point $\alpha=d$, the power-law form collapses and changes into the logarithmic form:
\begin{eqnarray}
    K\tau_{\mathrm{s}}^{\alpha=d} \sim (\ln L)^{-2}.
\end{eqnarray}

\end{widetext}

\section{The scaling law of the total evolution time}\label{app:4}

For the zero-momentum collective-spin XYZ Hamiltonian \eqref{H-ZM}, the evolution time to create the GHZ-like state is given by \cite{Zhang2024}
\begin{eqnarray}\label{tc}
    K t_\mathrm{c} \simeq \frac{6\ln N}{\lambda K \tau N^2}.
\end{eqnarray}
Since the scaling law of the pulse separation has been obtained as $K\tau\sim L^{-\mu}$,  the remaining step to estimate the scaling behavior of the evolution time $t_\mathrm{c}$ is to determine the parameter $\lambda$, which characterizes the average three-body interaction strength. Using the similar method to the one in Appendix~\ref{app:3}, we can approximately convert Eq.~\eqref{expression-lambda} into an integral
\begin{eqnarray}
     \lambda^{d=1} &\approx& \frac{1}{N^2}\left(2\int_{1}^{L/2}\mathrm{d}r\frac{1}{r^\alpha}\right)^2,\nonumber\\
     \lambda^{d=2} &\approx& \frac{1}{N^2}\left(\int_{1}^{L/2}2\pi r\mathrm{d}r\frac{1}{r^\alpha}\right)^2.
\end{eqnarray}
The scaling law of $\lambda$ can be directly obtained from the above equation, which gives
\begin{eqnarray}
     \lambda \sim \left\{
     \begin{aligned}
     & L^{-2\alpha} \quad &\alpha<d, \\
     & L^{-2d} \left(\ln L\right)^2 &\alpha=d,\\
     & L^{-2d} &\alpha>d.
     \end{aligned}
     \right.
\end{eqnarray}
Substituting this result into Eq.~\eqref{tc}, along with the result \eqref{mu}, we immediately find the scaling law of the evolution time: $t_{\mathrm{c}}\sim L^{-\nu}\ln L$ with
\begin{eqnarray}
    \nu = \left\{
		\begin{aligned}
		& d-\alpha \quad &\alpha<d+2, \\
		& -2 \quad &\alpha\geq d+2.
		\end{aligned}
		\right.
\end{eqnarray}

\section{The power-law exponent of the suitable pulse separation}\label{app:5}

\begin{figure}[th]
\includegraphics[width=\columnwidth]{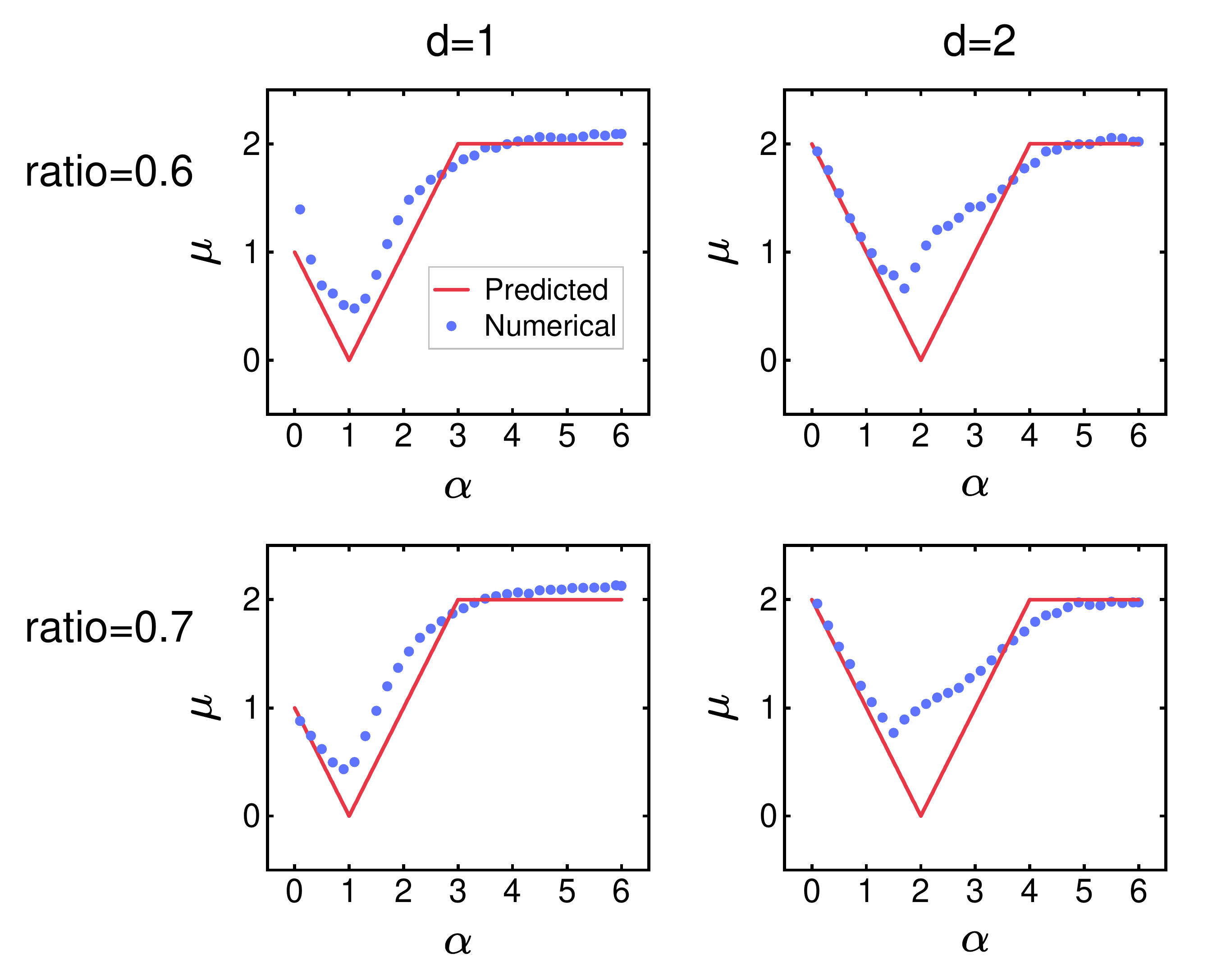}
\caption{The power-law exponent $\mu$ of the suitable pulse separation $\tau_{\mathrm{s}}$, approximately obeying $\tau_{\mathrm{s}}\sim L^{-\mu}$ with $\mu$ relying on $\alpha$. Numerical results are fitted using different critical ratios $F_{\mathrm{Q}}/F_{\mathrm{Q}}^{\mathrm{eff}} = 0.6,\,0.7$.}
\label{fig:6}
\end{figure}

In Fig.~\ref{fig:4}, we have verified the predicted power-law exponent $\mu$ by fixing $F_{\mathrm{Q}}/F_{\mathrm{Q}}^{\mathrm{eff}}=0.8$ and numerically fitting it. Fitted results using other ratios $F_{\mathrm{Q}}/F_{\mathrm{Q}}^{\mathrm{eff}}=0.6,\,0.7$ are provided in Fig.~\ref{fig:6}. It is obvious that different ratios give very similar results, showing that the scaling behavior does not rely on the artificial choice of the critical ratio.

\bibliography{alphaGHZlike}

\end{document}